\documentclass[journal]{IEEEtran}
\IEEEoverridecommandlockouts
\usepackage[dvipsnames]{xcolor}
\usepackage{cite}
\usepackage{amsmath,amssymb,amsfonts}
\usepackage{graphicx, tikz}
\usetikzlibrary{calc}
\usepackage{textcomp, multicol}
\usepackage{breqn}
\usepackage{amsmath}
\usepackage{todonotes}
\usepackage{bm}
\usepackage{bbm}
\usepackage{comment}
\usepackage{algorithm}
\usepackage{algpseudocode}
\usepackage{array}
\usepackage{threeparttable}
\usepackage{siunitx}
\usepackage{kantlipsum}
\usepackage{tabularx,booktabs}
\usepackage{pifont}% http://ctan.org/pkg/pifont
\newcommand{\cmark}{\ding{51}}%

\newcommand{\xmark}{\ding{55}}%
\newcolumntype{C}{>{\centering\arraybackslash}X} 
\def\BibTeX{{\rm B\kern-.05em{\sc i\kern-.025em b}\kern-.08em
T\kern-.1667em\lower.7ex\hbox{E}\kern-.125emX}}
\begin{document}
%\newgeometry{left=1.62cm,right=1.62cm,top=2.0cm} % Change to 3cm bottom (temp)
\thispagestyle{empty}
\title{Deep Reinforcement Learning Enhanced Rate-Splitting Multiple Access for Interference Mitigation}

\author{Osman~Nuri~Irkıçatal, Elif~Tugce~Ceran, Melda~Yuksel,~\IEEEmembership{Senior Member,~IEEE}
\thanks{The material in this paper was presented
in part at the IEEE Global Communications Conference,
Kuala Lumpur, Malaysia, December 2023 \cite{irkicatal2023deep}.}
\thanks{O. N. Irkıçatal is with both the Department of Electrical and Electronics Engineering, Middle East Technical University, Ankara, 06800, Turkey and with Aselsan Inc. Ankara, 06800, Turkey, e-mail: oirkicatal@aselsan.com.tr.}
\thanks{E. T. Ceran and M. Yuksel are with the Department of Electrical and Electronics Engineering, Middle East Technical University, Ankara, 06800, Turkey, e-mail: \{elifce, ymelda\}@metu.edu.tr.}
}

\maketitle

\begin{abstract}

This study explores the application of the rate-splitting multiple access (RSMA) technique, vital for interference mitigation in modern communication systems. It investigates the use of precoding methods in RSMA, especially in complex multiple-antenna interference channels, employing deep reinforcement learning. The aim is to optimize precoders and power allocation for common and private data streams involving multiple decision-makers. A multi-agent deep deterministic policy gradient (MADDPG) framework is employed to address this complexity, where decentralized agents collectively learn to optimize actions in a continuous policy space. We also explore the challenges posed by imperfect channel side information at the transmitter. Additionally, decoding order estimation is addressed to determine the optimal decoding sequence for common and private data sequences. Simulation results demonstrate the effectiveness of the proposed RSMA method based on MADDPG, achieving the upper bound in single-antenna scenarios and closely approaching theoretical limits in multi-antenna scenarios. Comparative analysis shows superiority over other techniques such as MADDPG without rate-splitting, maximal ratio transmission (MRT), zero-forcing (ZF), and leakage-based precoding methods. These findings highlight the potential of deep reinforcement learning-driven RSMA in reducing interference and enhancing system performance in communication systems.
\end{abstract}


\section{Introduction}

The imperative for high data rates and dependable connections in 6G networks is a direct response to the increasing demands of data-intensive applications and services\cite{jiang2021road}. Building on the groundwork laid by 5G, the move to 6G is driven by cutting-edge technologies like augmented reality, virtual reality, the internet of things (IoT), high-definition video streaming, remote healthcare, and autonomous systems\cite{lu20206g}. These applications not only require faster data transfers but also continuous and reliable connections, demanding advancements beyond the capabilities of current networks.

Achieving high data rates in 6G networks faces a significant challenge due to interference, especially in ultra-dense deployments of small cells and connected devices, where devices are in close proximity of each other. While ultra-densification boosts network capacity, it complicates interference management \cite{siddiqui2023urllc}. To address interference, advanced techniques are crucial, including sophisticated signal processing, spectrum-sharing strategies, cognitive radio systems, artificial intelligence algorithms, and innovative modulation schemes \cite{wei2023integrated}, \cite{salahdine20235g}. %These methods collectively comprise a robust toolkit for mitigating interference and optimizing spectrum usage in ultra-dense 6G networks. The shift from 5G to 6G is driven by the need to support diverse data-intensive applications. This evolution involves not only upgrading technologies but also addressing challenges posed by ultra-densification, essential to meet the growing demands of a connected world\cite{salahdine20235g}. 

Among these sophisticated techniques, rate-splitting is among the most robust and effective to mitigate interference. This method was first proposed in 1981 \cite{han1981new}. For example, in a two-user interference channel, there are two transmitters and two receivers, both of whom are only interested in their corresponding messages. In an interference channel, the regular approach is to treat the other user's message as noise. However, this other user's message is not noise, it has a structure and can be decoded if desired. To create a balance between treating unwanted messages as noise and fully decoding them, other user's messages can be partially decoded, and the remaining part can be treated as noise. This balance is achieved by constructing the messages in common and private parts. Rate-splitting uses successive interference cancellation (SIC) at the receivers to enable a step-by-step decoding process for all common messages and for individual private messages. This adaptability makes the rate-splitting method unique and achieves the largest rate region in interference channels. Later, it was discovered that multi-antenna broadcast channels require as much interference mitigation as interference channels, and rate-splitting is quite useful in ultra-dense 6G networks \cite{clerckx2023primer}, in integrated satellite and aerial networks \cite{lin2021supporting} or in reconfigurable intelligent surface-aided communication systems \cite{wu2023deep}. %Ensuring receivers decode common messages before extracting private information, rate-splitting offers a new strategy to optimize communication in interference. 
Since its introduction, rate-splitting has been a focal point in information theory, shaping advances in interference management. Researchers explore its applications, complexities, performance boundaries, and practical implementations, highlighting its significance  in enhancing communication efficiency in high-interference environments in both interference and multi-antenna broadcast channels \cite{dizdar2021rate}.

%This quality is valuable for upcoming communication systems, especially in ultra-dense scenarios, sustaining high data rates and reliable connectivity. RSMA's utility extends to various contexts, including IoT communications\cite{lin2021supporting}, wireless sensor networks, and scenarios requiring secure and efficient data transmission in interference-prone environments, making it promising for creating smooth, high-rate, and interference-resistant communication systems across diverse applications and network environments.

While RSMA offers high flexibility, tackling the non-convex optimization challenges for RSMA is particularly challenging \cite{mao2022rate}. Algorithms like the iterative weighted minimum mean square error (WMMSE) algorithm \cite{yalcin2018downlink} can be utilized to address the traditional non-convex sum-rate maximization problem. However, they come with significant computational complexity due to the need for high-dimensional matrix inversions and an excessive number of iterations. This challenge becomes more severe  in scenarios with multiple users, unknown or complex channel characteristics,  or increased rate-splitting layers. Consequently, finding the best rate-splitting combination becomes computationally challenging, necessitating sophisticated algorithms and robust computational resources, such as MATLAB CVX tools \cite{li2023sum}. 

%Devising strategies to address this complexity is crucial, significantly impacting the practical implementation and effectiveness of RSMA in real-world communication systems.
Another practical approach is to use lower-complexity heuristic methods, such as zero forcing (ZF) or maximum ratio transmission (MRT) \cite{du2021cell}. Although these algorithms are computationally efficient, their performance is often limited because ZF disregards residual interference, while MRT focuses solely on maximizing signal strength without considering interference management or user-specific channel conditions. As a result, they do not achieve satisfactory system performance in our scenario.

Deep reinforcement learning (DRL) is pivotal for addressing these challenges in RSMA \cite{arulkumaran2017deep}. RSMA's complexity requires advanced navigation in high-dimensional spaces, and DRL, within a Markov decision process (MDP), refines rate-splitting in dynamic interference scenarios, enhancing communication efficiency. %Its adaptability in handling high-dimensional spaces is valuable for addressing optimization challenges, crucial in RSMA, where navigating expansive solution spaces is crucial. Reinforcement learning, especially DRL, is increasingly used in 5G and 6G to address communication challenges due to its adaptability. Unlike traditional machine learning, RL learns through interactions, suitable for scenarios with unclear system behavior. 
Reinforcement learning excels in handling uncertainties, optimizing resource allocation, and precoder optimization \cite{dahrouj2021overview}. In terms of computational complexity, the main advantage of DRL method is that  it delegates the majority of the complex computations to the training phase, leaving only simple forward calculations to be performed during the implementation phase. Unlike standard deep learning (DL) algorithms, DRL algorithms can adapt to changing channel conditions and does not require a fixed, large dataset for training. Furthermore, multi-agent deep reinforcement learning (MADRL) employs decentralized decision-making and aligns with distributed communication systems such as interference channels considered in this paper, and effectively manages the present nonlinear dynamics. %RL's capacity for exploring new strategies is valuable in the evolving landscape of 6G. Despite other machine learning approaches, RL's unique features make it a valuable tool for optimizing communication networks for performance, efficiency, and adaptability.

% \begin{figure}[!htb]
%     \centering
%     \includegraphics[width=90mm]{figures/RSMA_illustration.PNG}
%     \caption{Illustration of a typical RSMA scenario \cite{mao2018rate}.}
%     \label{fig1:RSMA_illustration}
% \end{figure}

% \begin{figure}[!htb]
%     \centering
%     \includegraphics[width=90mm]{figures/RL_illustration.PNG}
%     \caption{The basic reinforcement learning scenario
%     \cite{amiri2018machine}.}
%     \label{fig1:RL_basic}
% \end{figure}

\subsection{Related Literature}

Unlike the recent literature on RSMA, which is for multi-antenna broadcast channels, in this paper we consider rate-splitting for the interference channel. Research on interference channels has been pivotal in information theory and communications, with seminal contributions shaping our understanding. %Shannon's influential work in the 1940s and 1950s, notably ``A Mathematical Theory of Communication'' \cite{shannon1948mathematical}, laid the theoretical foundation. Significant progress occurred in the 1970s and 1980s, with Cover and El Gamal's work \cite{el1980multiple} offering insights into multiple access channels and interference. Additionally, 
Han and Kobayashi's study in 1981 \cite{han1981new} proposed rate-splitting as a unique method for interference channels, introducing it as an innovative technique. Rate-splitting divides transmitted messages into common and private layers, addressing interference challenges by balancing noise treatment and efficient decoding \cite{clerckx2023primer}. %introduced innovative strategies for data transmission over interference channels, achieving previously unattainable rates. These key works have greatly advanced our comprehension of interference in communication systems.

While the exact capacity region of interference channels remains unknown, the seminal work \cite{sato1981capacity} studies the Gaussian interference channel, providing insights into achievable communication rates under strong interference conditions. Similarly, \cite{etkin2008gaussian}, and \cite{karmakar2013capacity} provide upper bounds within approximately one bit of the true capacity region for single-antenna and multiple-antenna interference channels, respectively. %Despite the challenge of determining the exact capacity region, these upper bounds serve as crucial benchmarks, indicating potential limits of information transmission in both single and multiple antenna interference scenarios. These findings significantly contribute to our understanding, guiding further exploration and development of communication strategies in high-interference environments. Another study by Sato in 1981 \cite{sato1981capacity} focuses on Gaussian channels, mathematically modeling noise and presenting the specific mathematical form of the Gaussian interference channel after normalization. The goal is to define theoretical limits on data transmission rates in the Gaussian Interference Channel, providing insights into achievable communication rates under strong interference conditions.

%Han and Kobayashi \cite{han1981new} proposed rate-splitting as a unique method for interference channels, introducing it as an innovative technique. Rate-splitting divides transmitted messages into common and private layers, addressing interference challenges by balancing noise treatment and efficient decoding \cite{clerckx2023primer}. Ensuring receivers decode common messages before extracting private information, rate-splitting offers a new strategy to optimize communication in interference. Since its introduction, rate-splitting has been a focal point in information theory, shaping advancements in interference management within modern communication networks \cite{wu2023deep}. Researchers explore its applications, complexities, performance boundaries, and practical implementations, highlighting its significance \cite{dizdar2021rate} in enhancing communication efficiency in high-interference environments.

The integration of DRL in communication systems introduces innovative approaches within a dynamic landscape. Utilizing the MDP framework, \cite{rahmani2022multi} employ a MADRL scheme based on the multi-agent deep deterministic policy gradient (MADDPG) algorithm, focusing on optimal precoders for the downlink. While earlier research \cite{lee2021multi} and \cite{lee2020deep} explore precoding challenges in multi-cell multi-user interference channels, they do not incorporate rate-splitting techniques. Studies like \cite{hieu2021optimal} and \cite{huang2022deep} apply DRL methods, specifically the proximal policy optimization (PPO) algorithm, to address power allocation issues in single-cell communications while incorporating rate-splitting. Additionally, \cite{naser2023deep} explores resource management and interference handling in sensing, energy harvesting, and communication functions using the trust region policy optimization (TRPO) approach. Recent advancements, exemplified by \cite{mismar2019deep}, leverage Q-learning, particularly the deep Q-Network (DQN), to maximize signal-to-interference-noise ratio (SINR) in multi-access orthogonal frequency division multiplexing (OFDM) networks. This highlights the potential of DRL in enhancing network performance without relying solely on rate-splitting. A summary of the existing works in the literature is provided in Table \ref{RelatedWork}.

%{\color{blue}Studies such as \cite{hua2023learning}, \cite{huang2020reconfigurable}, and \cite{wu2023deep} investigate the integration of RSMA and reconfigurable intelligent surface (RIS) techniques in next-generation networks. They target maximizing sum rates in RSMA IoT networks, enhancing resilience against imperfect channel information in terahertz multi-user Multiple Input Multiple Output (MIMO) systems, and optimizing resource efficiency in cellular networks through joint base stations (BS) and RIS design. Using DRL, hybrid data-model driven schemes, and novel optimization frameworks, these studies aim to enhance spectral efficiency while considering various network metrics. A summary of the existing works in the literature is provided in Table \ref{RelatedWork}.  The key concept is to view coordination challenges as scenarios where multiple agents collaborate in interference channels. These agents need to learn and adjust, working together to enhance communication performance. Notably, this groundbreaking approach introduces a new use of DRL. It addresses the intricate problems of power allocation and precoder design in interference channels, particularly incorporating rate-splitting techniques. This innovative approach represents a noteworthy progress, being the first instance of applying DRL to tackle power allocation and precoder issues in interference channels with rate-splitting complexities. This opens the door for more effective and adaptive interference management strategies.}

%%{|l|l|l|l|l|l|} 
\setlength{\extrarowheight}{1pt}
\begin{table*}
\caption{Some existing works that employ DRL in multi-user communication}
\label{RelatedWork}
\begin{tabularx}{\textwidth}{@{} l *{10}{C} c @{}}
\toprule
\textbf{Papers} & \textbf{DRL Method} & \textbf{Investigated Channel}  &  \textbf{Optimization Problem}  & \textbf{Inclusion of RSMA}  & \textbf{Cell/Antenna Configuration} \\
\midrule
\cite{rahmani2022multi}      & MADQN, MADDQN, MAD3QN     & Broadcast channel  &  Pilot contamination  & {\xmark}   & Cell-free massive MIMO      \\ 
\cite{lee2021multi,lee2020deep} & {MADDPG/DDPG}  & Interference channel     & {Precoder}  & {\xmark}   & {Multi-user Multi-cell, single-cell MISO}    \\ 
\cite{hieu2021optimal,huang2022deep} & {PPO}     & Broadcast channel   & {Precoder and power allocation coefficient} & {\cmark}     & Multi-user single-cell SISO
\\ 
\cite{naser2023deep}       & {TRPO}    &  Broadcast channel    & {Energy harvesting, sensing and communication capabilities}  &  {\cmark}  & Multi-user single-cell MISO            
\\ 
\cite{mismar2019deep}   & {DQL}     & Interference channel  & {Beamformer and power}   & {\xmark}      & Multi-user multi-cell MISO             
\\
\cite{vaezi2023deep}   & {DRL}     & Interference channel   & {Beamformer and power}   & {\xmark}      & Single-user multi-cell MISO              
\\
\cite{diamanti2023energy}   & {DQL}     & Broadcast channel   & {Energy efficiency}   & {\xmark}      & Multi-user single-cell MISO              
\\
\cite{muy2021energy}  & {MADQL}     & Broadcast channel   & Energy efficiency and power   & {\xmark}      & Multi-user single-cell MISO        \\
\cite{hua2023learning,huang2020reconfigurable,wu2023deep}  & {DL/DDPG}     & Broadcast channel   & {Precoder and power}   & {\cmark/\xmark}     & Multi-user single-cell MIMO             
\\
This work  & {MADDPG}     & Interference channel   & {Precoder and power allocation coefficient}   & {\cmark}    & {multi-user SISO, multi-user MISO, multi-user MIMO}    
\\
\bottomrule
\end{tabularx}
\end{table*}

\subsection{Contributions and Novelties}
%%%%%%%%%%%%%%%%%%%%%%%%%%%%%%%%%%%%%%%%%%%%%%%%%%%%%%%%%%%%%%%%%%%%%%%%%%%%%%%%%%%%%%%%%%%%%%%%%%%%

Existing research on RSMA and the broadcast channel (BC) within wireless communication networks showcases diverse approaches, some integrating learning algorithms while others rely on conventional optimization methods. These studies focus on enhancing efficiency and resource allocation but often overlook the learning aspects associated with decoding orders and channel estimation errors.

Numerous investigations \cite{li2023sum}, \cite{tang2023energy} delve into RSMA for the downlink broadcast channel, addressing resource allocation, power control, and spectral efficiency without employing learning algorithms. Instead, they utilize traditional optimization techniques, heuristic approaches, or game theory principles to optimize system performance.

Conversely, a subset of research within RSMA and downlink broadcast channel domains, \cite{hieu2021optimal}, \cite{huang2022deep}, \cite{naser2023deep}, leverage learning algorithms; %particularly DRL or other machine learning (ML) methods. These studies explore optimal resource allocation, power distribution, and rate assignment through adaptable AI-based approaches, contributing to improved performance in communication networks.
however, they are not directly applicable to interference channels. Moreover, they neglect learning decoding orders or addressing channel estimation errors. %These elements, critical in optimizing network efficiency and reliability, are yet to be thoroughly examined within the context of RSMA and BC paradigms. Therefore, while current research focuses on resource allocation and performance enhancement, there remains untapped potential in exploring learning mechanisms for decoding and mitigating channel estimation errors in interference channels. 
The key contributions of this work are summarized as follows:

\begin{itemize}   
    \item The work introduces a novel MADDPG algorithm customized for optimizing precoding and power allocation coefficients in multiple antenna interference channels employing rate-splitting strategies.

    \item The algorithm's framework allows for centralized learning while enabling decentralized execution, which contributes a decentralized and scalable framework for interference management without the need for constant coordination from a central entity.

    \item The work compares the performance of the proposed MADDPG algorithm against existing baseline schemes and upper bounds. It showcases the superiority of MADDPG with rate splitting. %, demonstrating optimal outcomes particularly in scenarios with multiple antennas at base stations and single antenna cases.
    
    \item The work investigates the impact of channel estimation errors, and incorporate optimal decoding order selection for common and private messages into the learning algorithm. These steps enhance the algorithm's robustness and broaden its scope of application.

    %\item The study provides a comprehensive analysis of RSMA, outlining its intricacies and challenges, particularly in managing interference and decoding strategies for multiple messages. It introduces and explores the utilization of DRL within RSMA networks, showcasing its potential to optimize resource allocation, address interference, and manage complex decoding strategies efficiently.

\end{itemize}
 
The remaining sections of this paper are structured as follows. We outline the system model for RSMA in Section~\ref{sec:system_model} and elaborate on the MADDPG tailored to our system model in Section~\ref{sec:MADDPG}. Section~\ref{sec:bench} introduces the benchmark schemes used for comparison. In Section~\ref{sec:sim}, we present the simulation results. Lastly, Section~\ref{sec:conclusion} includes our conclusions and outlines areas for future work.

\section{System Model}
\label{sec:system_model}

The system model considers a multiple input multiple output (MIMO) interference channel comprising two base stations ($BS$) each having $M_1$ and $M_2$ antennas and two users each having $N_{1}$ and $N_{2}$ antennas, respectively paired with each $BS_i$, $i=1,2$. The $BS_i$ sends the message $S_i$ to user equipment, $UE_i$. The number of messages that can be transmitted in RSMA, ${Q_i}$, is restricted by the antenna configurations and is written as
\begin{equation}\label{eq:message_number}
{Q_i} = \min(M_i,N_i).
\end{equation}

In rate-splitting, ${S}_{i}$ is split into a common and a private part; i.e., $S_{i}^{c}$ and $S_{i}^{p}$. The $Q_{i}$ common and private messages at each $BS_i$ are independently encoded into streams $\bm{b}_{ic}$ and $\bm{b}_{ip}$ where ${\bm{b}_{ic}},{\bm{b}_{ip}} \in \mathbb{C}^{Q_{i} \times 1}$ and respectively precoded with ${\bm{W}_{ic}}$ and ${\bm{W}_{ip}}$, where ${\bm{W}_{ic}}$ and ${\bm{W}_{ip}} \in \mathbb{C}^{M_i \times Q_i}$. All messages $\bm{b}_{1c}$, $\bm{b}_{1p}$, $\bm{b}_{2c}$ and $\bm{b}_{2p}$ are independent from each other, and $\mathbb{E}\{\bm{b}_{in}\bm{b}_{in}^\ast\} = \bm{I}_{Q_i}$, $i=1,2$, $n=c,p$, where $\bm{I}_{Q_i}$ is the identity matrix of size $Q_i \times Q_i$. Then, the transmitted  signal of $BS_i$, ${\bm{x}_{i}} \in \mathbb{C}^{M_i \times 1}$ where $j \neq i$ and  ${i,j \in \{1,2\}}$, is defined as 
\begin{equation}\label{eq:x_i_mimo}
{\bm{x}}_i = {\bm{W}_{ic}}\bm{b}_{ic} + {\bm{W}_{ip}}\bm{b}_{ip}.
\end{equation}
To satisfy the power constraints at each one of the transmitters, we assume $|{\bm{w}}_{ikc}|^2 +|{\bm{w}}_{ikp}|^2 \leq P_{ik}$, where $\sum_{k=1}^{Q_i} P_{ik} = P_i$ is the total power of $BS_i$,  and $\bm{w}_{ikc}$ and  $\bm{w}_{ikp}$ represent the $k$th column of $\bm{W}_{ic}$ and  $\bm{W}_{ip}$ respectively, $k = 1,\ldots,Q_i$ \footnote{The values the variable $k$ takes depend on the value $i$. However, to keep the notation simple, we slightly abuse the notation and do not have an $i$ index for $k$.}. Moreover, we define 
\begin{eqnarray}
\bm{P}_{ic} &=& \left[P_{i1c}, \ldots, P_{iQ_ic}\right]^T= \left[|\bm{w}_{i1c}|^2, \ldots, |\bm{w}_{iQ_ic}|^2\right]^T\\
\bm{P}_{ip} &=& \left[P_{i1p}, \ldots, P_{iQ_ip}\right]^T= \left[|\bm{w}_{i1p}|^2, \ldots, |\bm{w}_{iQ_ip}|^2\right]^T
\end{eqnarray}
where $T$ denotes the transpose operation.

The received signal at user $UE_i$ is then written as 
\begin{align}
\bm{y_{i}} &= {\bm{H}}_{i}{\bm{x}}_{i} + {\bm{G}}_{j}{\bm{x}}_{j} + {\bm{n_{i}}}. 
\end{align}
Here $j$ indicates the index of the interfering signal, $j = 1,2$ and $j \neq i$. The channel gain between $BS_i$ and ${UE}_i$ is indicated as ${\bm{H}_{i}} \in \mathbb{C}^{N_{i} \times M_i}$. Similarly, the channel gain between $BS_j$ and ${UE}_i$ is  ${\bm{G}_{j}} \in \mathbb{C}^{N_{i} \times M_j}$. The entries in ${\bm{H}}_i$ and ${\bm{G}}_j$ are independent and identically distributed and complex valued random variables. The transmitters $BS_i$ are informed about their outgoing channel gains ${\bm{H}}_i$ and ${\bm{G}}_i$, while the receivers ${UE}_i$ know only their incoming channel gains ${\bm{H}}_i$ and ${\bm{G}}_j$  The noise term at ${UE}_i$ is denoted with $\bm{n}_i$. It is circularly symmetric complex Gaussian with mean zero and variance ${\sigma^2_{n,i}}{{\bm{I}_N}_{i}}$; i.e., ${\bm{n}_{i}} \in \mathcal{CN}\big(0,{\sigma^2_{n,i}}{\bm{I}_N}_{i}\big) $ . Also $\bm{n}_1$ and $\bm{n}_2$ are independent from each other. For simplicity, we will take $ {\sigma^2_{n,i}} = N_0 $. %In the following subsection rate expressions for rate-splitting are explained. 

In order to be able to write the achievable rates with RSMA, the decoding order for $\bm{b}_{ic}$ and $\bm{b}_{ip}$ must be determined in both users \cite{mao2018rate}. Moreover, to attain the best possible achievable rates, one has to consider all possible choices of these decoding orders. For the particular system model we study, we take 2 different decoding orders into consideration for each one of the users. Namely, ${UE}_i$ either decodes in the order (a) or (b) in \eqref{decoding order}.
\begin{align} 
(a) ~ \bm{b}_{jc} \rightarrow \bm{b}_{ic} \rightarrow \bm{b}_{ip}, \quad
(b) ~ \bm{b}_{ic} \rightarrow \bm{b}_{jc} \rightarrow \bm{b}_{ip}.  \label{decoding order}
\end{align}
For example, when $UE_1$ decodes according to the order given in (a) and $UE_2$ decodes according to (b), the achievable rates at $UE_1$ are as follows:
\begin{eqnarray}
R_{2c}^{1} &=& \log_2\det\left({I_{N_1}} + {\bm{G}_{2}}{\bm{W}_{2c}}{\bm{W}^H_{2c}}{\bm{G}^H_{2}} r_{1,2c}^{-1} \right),\label{eq:R2c1} \\
R_{1c}^{1} &=& \log_2\det\left({I_{N_1}} + {\bm{H}_{1}}{\bm{W}_{1c}}{\bm{W}^H_{1c}}{\bm{H}^H_{1}}r_{1,1c}^{-1}\right),\label{eq:R1c1}  \\
R_{1p} &=& \log_2\det\left({I_{N_1}} + {\bm{H}_{1}}{\bm{W}_{1p}}{\bm{W}^H_{1p}}{\bm{H}^H_{1}} r_{1,1p}^{-1} \right.),\label{eq:R1p} 
\end{eqnarray}
where
\begin{eqnarray*}
r_{1,2c} &=& {\sum_{n=\{c,p\}}{{{\bm{H}_{1}}}{\bm{W}_{1n}}{{\bm{W}^{H}_{1n}}}{\bm{H}^{H}_{1}}}} +  {{{{\bm{G}_{2}}{\bm{W}_{{2p}}}{\bm{W}^{H}_{{2p}}}{\bm{G}^{H}_{2}}}}} \nonumber \\
&&\:{+}{{N_{0}I_{N_1}}}, \\
r_{1,1c} &=& { {{{\bm{H}_{1}}}{\bm{W}_{1p}}{{\bm{W}^H_{1p}}}{\bm{H}^H_{1}}}}  +  {{{{{\bm{G}_{2}}}{\bm{W}_{2p}}{{\bm{W}^H_{2p}}}{\bm{G}^H_{2}}}}}  +{{N_{0}I_{N_1}}}, \\
r_{1,1p} &=&  {{{\bm{G}_{2}}}{\bm{W}_{2p}}{{\bm{W}^H_{2p}}}{\bm{G}^H_{2}}} + N_{0}I_{N_1}.
\end{eqnarray*}
Similarly, the achievable rates at $UE_{2}$ are as follows:
\begin{eqnarray} 
R_{2c}^{2} &=& \log_2\det\left({I_{N_2}} +  {\bm{H}_{2}}{\bm{W}_{2c}}{\bm{W}^H_{2c}}{\bm{H}^H_{2}}r_{2,2c}^{-1} \right),\label{eq:R2c2}\\
R_{1c}^{2} &=& \log_2\det\left({I_{N_2}} + {\bm{G}_{1}}{\bm{W}_{1c}}{\bm{W}^H_{1c}}{\bm{G}^H_{1}}r_{2,1c}^{-1} \right), \label{eq:R1c2} \\
R_{2p} &=& \log_2\det\left({I_{N_2}} + {\bm{H}_{2}}{\bm{W}_{2p}}{\bm{W}^H_{2p}}{\bm{H}^H_{2}}  r_{2,2p}^{-1} \right), \label{eq:R2p} 
\end{eqnarray}
where
\begin{eqnarray*}
r_{2,2c} &=&  \sum_{n=\{c,p\}}\bm{G}_{1}\bm{W}_{1n}\bm{W}^H_{1n}\bm{G}^H_{1} + \bm{H}_{2}\bm{W}_{2p}\bm{W}^H_{2p}\bm{H}^H_{2} \nonumber \\
&&\:{+} N_{0}I_{N_2}, \\
r_{2,1c} &=&  \bm{G}_{1}\bm{W}_{1p}\bm{W}^H_{1p}\bm{G}^H_{1}+ \bm{H}_{2}\bm{W}_{2p}\bm{W}^H_{2p}\bm{H}^H_{2} +N_{0}I_{N_2},\\
r_{2,2p} &=& \bm{G}_{1}\bm{W}_{1p}\bm{W}^H_{1p}\bm{G}^H_{1} + N_{0}I_{N_2}.
\end{eqnarray*}
Since common messages should be decoded at both receivers, common message rates are actually limited with the minimum of \eqref{eq:R2c1} and \eqref{eq:R2c2} and of \eqref{eq:R1c1} and \eqref{eq:R1c2}. Thus, we define $R_{1c}$ and $R_{2c}$ as
\begin{align}
R_{1c} &= \min (R_{1c}^{1},R_{1c}^{2}) \label{eq:R1cmin}\\
R_{2c} &= \min (R_{2c}^{1},R_{2c}^{2}). \label{eq:R2cmin} \end{align} Then, the rate for ${UE}_i$, $i=1,2$, can be calculated as
\begin{align} R_i = R_{ic}+R_{ip}. \label{eq:R1sum} \end{align}
Note that one can write 4 different sets of achievable rates as in \eqref{eq:R2c1}-\eqref{eq:R2p}, considering different combinations of decoding orders listed in (a) and (b) in \eqref{decoding order}. Then, for a given $\beta \in [0,1]$, for user rates, and a given decoding order, the objective is to maximize 
% \begin{subequations}
% \begin{align}
% \max_{\bm{W}_{ic}, \bm{W}_{ip}, \bm{P}_{ic}, \bm{P}_{ip}, i = 1,2} &\qquad \beta R_1 + (1-\beta) R_2 \label{eq:maxrateobj} \\
% \text{s.t. }
% &\qquad  |\bm{w}_{ikc}|^2+|\bm{w}_{ikp}|^2 \leq P_{ik}, 
% \label{eq:Piconstant} \\ &  \quad \quad \quad \quad \quad i=1,2,  \quad k=1,2,..,Q_i \nonumber
% \\ & \qquad \quad    \sum_{k=1}^{Q_{i}} P_{ik} = P_{i}, \quad i = 1,2.\label{eq:Pisum}
% \end{align}
% \end{subequations}
\begin{subequations}
\begin{align}
\max_{\bm{W}_{ic}, \bm{W}_{ip},  ~i = 1,2} &\qquad \beta R_1 + (1-\beta) R_2 \label{eq:maxrateobj} \\
\text{s.t. }
&\qquad  \sum_{k=1}^{Q_{i}} \left(|\bm{w}_{ikc}|^2+|\bm{w}_{ikp}|^2\right) \leq P_{i}, \quad i = 1,2.\label{eq:Pisum}
%&  \quad    \sum_{k=1}^{Q_{i}} |\bm{w}_{ikc}|^2+|\bm{w}_{ikp}|^2 \leq P_{i}, \quad i = 1,2.\label{eq:Pisum}
\end{align}
\end{subequations}

%MIMO RSMA outperforms SISO and MISO RSMA configurations in several ways. With multiple antennas at both ends, MIMO RSMA enables simultaneous transmission of multiple data streams enhancing spectral efficiency and offering higher data rates. It harnesses spatial diversity and multiplexing gains, effectively managing interference and improving reliability against channel fading. The system's adaptability and robustness to varying channel conditions further highlight its superiority, allowing for adaptive strategies and increased overall flexibility in wireless communication setups.

%However, its complexity escalates notably compared to SISO and MISO RSMA systems. The increased complexity primarily stems from the need to handle multiple antennas at both the transmitter and receiver ends, resulting in heightened signal processing demands and computational requirements. MIMO RSMA involves intricate spatial processing, necessitating sophisticated algorithms for beamforming, precoding, and decoding across multiple antenna elements. Moreover, managing interference and decoding orders becomes more challenging with the increased dimensionality and complexity introduced by multiple antennas. Therefore, learning based solution will be presented in the next section.

The complexity of the above optimization problem is quite high, when regular optimization tools are used. Therefore, in the next section, we will present the MADDPG with RSMA for MIMO interference channels. 

\section{MADDPG for Precoding and Power Allocation Coefficients Optimization}\label{sec:MADDPG}

In this section, we propose to use multi-agent deep reinforcement learning algorithm with decentralized policies and joint action optimization in order to solve the average sum-rate maximization problem defined in Section~\ref{sec:system_model}. Specifically, we adopt the MADDPG algorithm~\cite{lowe2017multi}, which is an extension of the well-known deep deterministic policy gradient (DDPG) algorithm \cite{lillicrap2015continuous} tailored specifically for multi-agent systems. It is a powerful algorithm that has been successfully applied to challenging tasks in signal processing and communication areas \cite{rahmani2022multi},\cite{albinsaid2021multi}.

% \begin{figure*}[tb]
%     \centering    
%     \includegraphics[width=\textwidth]{figures/SYSTEM_architecture_MIMO.png}
%     \caption{System architecture for MADDPG with RSMA for MIMO interference channels.}  \label{fig:mimo_arch}
% \end{figure*}

MADDPG employs centralized training and decentralized execution, where all agents share a common critic network to facilitate joint action optimization, and decentralized execution, where each agent independently executes its learned policy based on local observations. Specifically, the critic network in MADDPG takes as input not only the local observations and actions of an individual agent, but also the observations and actions of all other agents in the system. By doing so, the critic can learn a centralized value function that takes into account the joint actions of all agents. This centralized value function can then be used to train each agent's policy network.

On the other hand, during execution or deployment, each agent only has access to its own local observations and actions, namely, agent $i$ at $BS_i$ chooses the precoding matrices $\bm{W}_{ic}$, $\bm{W}_{ip}$ and power allocation coefficient vectors $\bm{P}_{ic}$ (and inherently $\bm{P}_{ip}$) based on local information characterized by $\bm{H}_{i}$ and $\bm{G}_{i}$ only. This is known as decentralized execution, as each agent acts independently based on its own observations and policies without relying on information from other agents. By decoupling the execution from the learning, MADDPG is able to handle complex multi-agent systems, where agents have limited or incomplete information about the system as a whole.

% \begin{figure*}[tb]
%     \centering
%     \includegraphics[scale = 0.3]{figures/MIMO_MADDPG.png}
%     \caption{MADDPG with RSMA algorithm structure for the MIMO case.}
%     \label{Figure:system_architecture}
% \end{figure*}

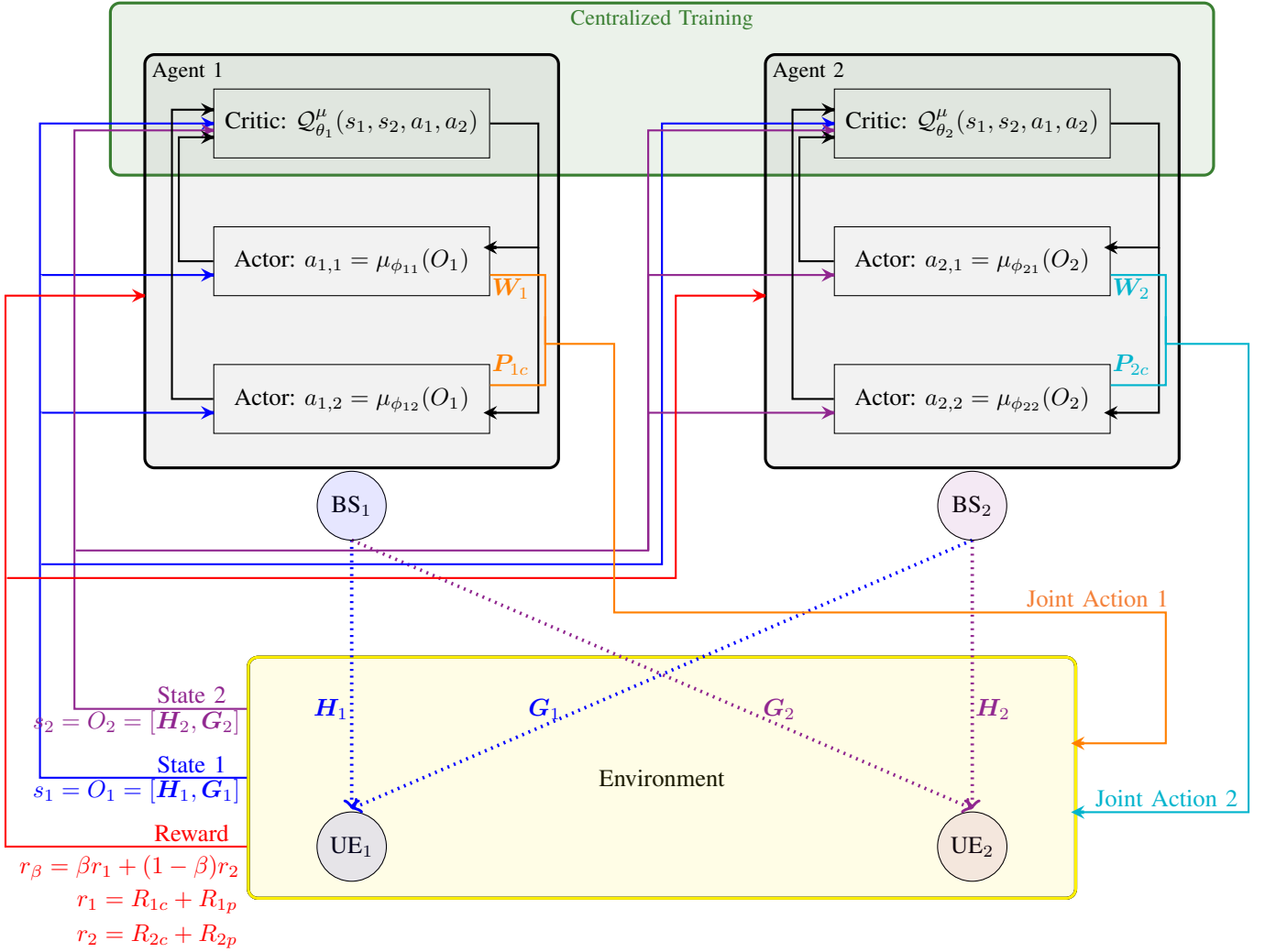
\begin{figure*}
\begin{multicols}{2}
\begin{centering}    
\begin{tikzpicture}[
    brect/.style={draw=OliveGreen, very thick, minimum width=16cm, minimum height=2.5cm, align=center}, % Centralized learning box
    rect/.style={draw, very thick, minimum width=12cm, minimum height=3.5cm, align=center}, % Environment box wider
    arect/.style={draw, very thick, minimum width=6cm, minimum height=6cm, align=center}, % Agent box
    innerrect/.style={draw, minimum width=4cm, minimum height=1cm, align=center}, % Inner rectangles
    arrow/.style={-, thick} % Style for arrows
]
%Centralized learning box
\node[brect, fill=OliveGreen, fill opacity=0.1, rounded corners] (centlearn) at (0, 6.5) {};
\node[anchor=north, font=\small] at (centlearn.north) {{\color{OliveGreen}Centralized Training}};

% Environment
\node[rect, rounded corners] (env) at (0, -3.5) {Environment};
\node[rect, fill=yellow, fill opacity=0.1, rounded corners] at (0, -3.5) {};
\node[rect, draw=yellow, rounded corners] at (0, -3.5) {};
\node[circle, draw, minimum size=1cm, align=center] (circle1) at (-4.5, -4.5) {UE$_1$};
\node[circle, draw, minimum size=1cm, align=center] (circle1) at (4.5, -4.5) {UE$_2$};
\node[circle, fill=blue, fill opacity=0.1,minimum size=1cm] at (-4.5, -4.5) {};
\node[circle, fill=Plum, fill opacity=0.1,minimum size=1cm] at (4.5, -4.5) {};

% Wireless Channel
\draw[arrow, draw= blue, dotted, ->, very thick, line width=0.5mm] (-4.5,-0.05) -- ++(0, -3.9);
\draw[arrow, draw= Plum,dotted, ->, very thick, line width=0.5mm] (-4.5,-0.05) -- ++(9, -3.9);
\draw[arrow, draw= Plum,dotted, ->, very thick, line width=0.5mm] (4.5,-0.05) -- ++(0, -3.9);
\draw[arrow, draw= blue,dotted, ->, very thick, line width=0.5mm] (4.5,-0.05) -- ++(-9, -3.9);
\node at (-4.8, -2.5) {\color{blue}$\bm{H}_1$};
\node at (4.8, -2.5) {\color{Plum}$\bm{H}_2$};
\node at (-1.7, -2.5) {\color{blue}$\bm{G}_1$};
\node at (1.7, -2.5) {\color{Plum}$\bm{G}_2$};

% Agent 1 (left)
\node[arect, fill=gray, fill opacity=0.1, rounded corners] (agent1) at (-4.5, 4) {};
\node[innerrect] (critic1) at (-4.5, 6) {Critic: $\mathcal{Q}_{\theta_1}^\mu(s_1,s_2,a_1,a_2)$};
\node[innerrect] (actor11) at (-4.5, 4) {Actor: $a_{1,1}=\mu_{\phi_{11}}(O_1)$};
\node[innerrect] (actor12) at (-4.5, 2) {Actor: $a_{1,2}=\mu_{\phi_{12}}(O_1)$};
\node[anchor=north west, font=\small] at (agent1.north west) {Agent 1};
\node[circle, draw, minimum size=1cm, align=center] (circle1) at (-4.5, 0.45) {BS$_1$};
\node[circle, fill=blue, fill opacity=0.1,minimum size=1cm] at (-4.5, 0.45) {};

% Agent 2 (right)
\node[arect, fill=gray, fill opacity=0.1, rounded corners] (agent2) at (4.5, 4) {};
\node[innerrect] (critic2) at (4.5, 6) {Critic: $\mathcal{Q}_{\theta_2}^\mu(s_1,s_2,a_1,a_2)$};
\node[innerrect] (actor21) at (4.5, 4) {Actor: $a_{2,1}=\mu_{\phi_{21}}(O_2)$};
\node[innerrect] (actor22) at (4.5, 2) {Actor: $a_{2,2}=\mu_{\phi_{22}}(O_2)$};
\node[anchor=north west, font=\small] at (agent2.north west) {Agent 2};
\node[circle, draw, minimum size=1cm, align=center] (circle1) at (4.5, 0.45) {BS$_2$};
\node[circle, fill=Plum, fill opacity=0.1,minimum size=1cm] at (4.5, 0.45) {};

% State arrows from environment to agent 1
\draw[arrow, draw=blue] ($(env.west) + (0, 0)$) -- ++(-3, 0) -- ++(0,9.5) node at ($(env.west) + (-0.8, 0.2)$) {\color{blue}State 1} |- ($(critic1.west)+(0,0)$);
\node at ($(env.west) + (-1.6, -0.2)$) {\color{blue}$s_1 = O_1 = [\bm{H}_1,\bm{G}_1]$};
\draw[->, >=stealth, draw=blue, line width=0.5mm] (-6.5,3.8) -- ++(0.01, 0);
\draw[->, >=stealth, draw=blue, line width=0.5mm] (-6.5,6) -- ++(0.01, 0);
\draw[arrow, draw= blue] (-9,1.8) |- ($(actor12.west)+(0,-0.2)$);
\draw[arrow, draw= blue] (-9,3.8) |- ($(actor11.west)+(0,-0.2)$);
\draw[arrow, draw=Plum] ($(env.west) + (0, 1)$) -- ++(-2.5, 0) -- ++(0,8.4) node at ($(env.west) + (-0.8, 1.2)$) {\color{Plum}State 2}|- ($(critic1.west)+(0,-0.1)$);
\draw[->, >=stealth, draw=Plum, line width=0.5mm] (-6.5,5.9) -- ++(0.01, 0);
\draw[->, >=stealth, draw=blue, line width=0.5mm] (-6.5,1.8) -- ++(0.01, 0);
\node at ($(env.west) + (-1.6, 0.8)$) {\color{Plum}$s_2 = O_2=[\bm{H}_2,\bm{G}_2]$};

% State arrows from environment to agent 2
\draw[arrow, draw=blue] (-9,-0.4) -- ++(9, 0) -- ++(0,6.1) |- (critic2.west);
\draw[arrow, draw=Plum] (-0.2,-0.2) -- ++(0,3.9) |- ($(actor21.west)+(0,-0.2)$);
\draw[arrow, draw=Plum] (-0.2,-0.2)  -- ++(0,1.9) |- ($(actor22.west)+(0,-0.2)$);
\draw[->, >=stealth, draw=blue, line width=0.5mm] (2.5,6) -- ++(0.01, 0);
\draw[->, >=stealth, draw=Plum, line width=0.5mm] (2.5,3.8) -- ++(0.01, 0);
\draw[->, >=stealth, draw=Plum, line width=0.5mm] (2.5,1.8) -- ++(0.01, 0);
\draw[arrow, draw=Plum] (-8.5,-0.2) -- ++(8.3, 0) -- ++(0,5.3) |- ($(critic2.west)+(0,-0.1)$);
\draw[->, >=stealth, draw=Plum, line width=0.5mm] (2.5,5.9) -- ++(0.01, 0);

% Reward arrow from environment to agents
\draw[arrow, draw=red] ($(env.west) + (0, -1)$) -- ++(-3.5, 0) -- ++(0,6.5) node at ($(env.west) + (-0.8, -0.8)$) {\color{red}Reward} |- ($(agent1.west)+(0,-0.5)$);
\node at ($(env.west) + (-1.7, -1.3)$) {\color{red}$r_{\beta}= \beta r_1 + (1-\beta)r_2$};
\node at ($(env.west) + (-1.3, -1.8)$) {\color{red}$r_{1}=R_{1c}+R_{1p}$};
\node at ($(env.west) + (-1.3, -2.3)$) {\color{red}$r_{2}=R_{2c}+R_{2p}$};
\draw[->, >=stealth, draw=red, line width=0.5mm] (-7.5,3.5) -- ++(0.01, 0);
\draw[arrow, draw=red] (-9.5,-0.6) -- ++(9.7, 0) -- ++(0,3.6) |- ($(agent2.west)+(0,-0.5)$);
\draw[->, >=stealth, draw=red, line width=0.5mm] (1.5,3.5) -- ++(0.01, 0);

% Output arrow from Actor 11 to Critic 1
\draw[arrow] (actor11.west) -- ++(-0.5, 0) -- ++(0, 1.8) -> ($(critic1.west)+(0,-0.2)$);
\draw[->, >=stealth, line width=0.5mm] (-6.5,5.8) -- ++(0.01, 0);
% Output arrow from Actor 12 to Critic 1
\draw[arrow] (actor12.west) -- ++(-0.6, 0) -- ++(0, 4.2) -> ($(critic1.west)+(0,0.2)$);
\draw[->, >=stealth, line width=0.5mm] (-6.5,6.2) -- ++(0.01, 0);
% Output arrow from Critic 1 to Actor 11
\draw[arrow] (critic1.east) -- ++(0.7, 0) -- ++(0, -1.8) -- ($(actor11.east)+(0,0.2)$);
\draw[-<, >=stealth, line width=0.5mm] (-2.4,4.2) -- ++(0.01, 0);
% Output arrow from Critic 1 to Actor 12
\draw[arrow] (critic1.east)  ++(0.7, 0)  ++(0, -1.8) --++(0, -2.4) -- ($(actor12.east)+(0,-0.2)$);
\draw[-<, >=stealth, line width=0.5mm] (-2.4,1.8) -- ++(0.01, 0);

% Output arrow from Actor 21 to Critic 2
\draw[arrow] (actor21.west) -- ++(-0.5, 0) -- ++(0, 1.8) -> ($(critic2.west)+(0,-0.2)$);
\draw[->, >=stealth, line width=0.5mm] (2.5,5.8) -- ++(0.01, 0);
% Output arrow from Actor 22 to Critic 2
\draw[arrow] (actor22.west) -- ++(-0.6, 0) -- ++(0, 4.2) -> ($(critic2.west)+(0,0.2)$);
\draw[->, >=stealth, line width=0.5mm] (2.5,6.2) -- ++(0.01, 0);
% Output arrow from Critic 2 to Actor 21
\draw[arrow] (critic2.east) -- ++(0.7, 0) -- ++(0, -1.8) -- ($(actor21.east)+(0,0.2)$);
\draw[-<, >=stealth, line width=0.5mm] (6.6,4.2) -- ++(0.01, 0);
% Output arrow from Critic 2 to Actor 22
\draw[arrow] (critic2.east)  ++(0.7, 0)  ++(0, -1.8) --++(0, -2.4) -- ($(actor22.east)+(0,-0.2)$);
\draw[-<, >=stealth, line width=0.5mm] (6.6,1.8) -- ++(0.01, 0);

% Action arrows from agent1 to environment
\draw[arrow, draw=orange] ($(actor11.east)+(0,-0.2)$) --++ (0.8,0) --++ (0,-1) --++ (1,0) --++ (0,-3.9) --++ (8,0) --++ (0,-1.8) node at ($(env.east) + (0.3, 2.6)$) {\color{Orange}Joint Action 1} |- ($(env.east)+(0,0.5)$);
\draw[arrow, draw=orange] ($(actor12.east)+(0,0.2)$) --++ (0.8,0) --++ (0,1);
\draw[-<, >=stealth, draw=orange, line width=0.5mm] ($(env.east)+(0.1,0.5)$) -- ++(0.01, 0);
\node at ($(actor11.east) + (0.3, -0.4)$) {\color{orange}$\bm{W}_{1}$};
\node at ($(actor12.east) + (0.3, 0.5)$) {\color{orange}$\bm{P}_{1c}$};

% Action arrows from agent2 to environment
\draw[arrow, draw=Turquoise] ($(actor21.east)+(0,-0.2)$) --++ (0.8,0) --++ (0,-1) --++ (1.2,0) --++ (0,-6)  node at ($(env.east) + (1.3, -0.3)$) {\color{Turquoise}Joint Action 2}|- ($(env.east)+(0,-0.5)$);
\draw[arrow, draw=Turquoise] ($(actor22.east)+(0,0.2)$) --++ (0.8,0) --++ (0,1);
\draw[-<, >=stealth, draw=Turquoise, line width=0.5mm] ($(env.east)+(0.1,-0.5)$) -- ++(0.01, 0);
\node at ($(actor21.east) + (0.3, -0.4)$) {\color{Turquoise}$\bm{W}_{2}$};
\node at ($(actor22.east) + (0.3, 0.5)$) {\color{Turquoise}$\bm{P}_{2c}$};

\end{tikzpicture}
\end{centering}
\end{multicols}
\caption{System and MADDPG with RSMA algorithm structure for MIMO interference channels. In the MADDPG algorithm, there is centralized training and distributed execution.}
\label{Figure:system_architecture}
\end{figure*}

Initially, we assume the optimal decoding order that maximizes the sum-rate and without loss of generality, we consider an environment with two agents each representing a user. The overall system architecture and the algorithm structure are summarized in Fig.~\ref{Figure:system_architecture}. Later, in Section \ref{sec:decoding}, we will extend this approach by determining the optimal decoding order using MADDP.

In this system, each agent~$i$ at $BS_i$ chooses a precoding matrix $\bm{W}_{i}=[\bm{W}_{ic} ~ \bm{W}_{ip}]$ and a power allocation coefficient vector $\bm{P}_{ic}$ based on the local observation $O_i=[\bm{H}_{i}~\bm{G}_i]$, $i\in \{1,2\}$. We construct actor networks parameterized by $\phi_{i1}$ for the precoding  matrix $\bm{W}_{i}=[\bm{W}_{ic} ~ \bm{W}_{ip}]$; and parameterized by $\phi_{i2}$ for the power allocation coefficients $\bm{P}_{ic}$, and a critic network to evaluate the performance of policies for each agent $i$. Please note that,  in the single antenna case, the actor network is solely responsible for selecting $\bm{P}_{ic}$. Here, ${P}_{ic}$ becomes a single power allocation coefficient for each agent, eliminating the need to determine $\bm{W}_{i}$. Let $\mu=\{\mu_{\phi_{11}}(O_1),\mu_{\phi_{12}}(O_1),\mu_{\phi_{21}}(O_2),\mu_{\phi_{22}}(O_2)\}$ denote the set of policies parameterized by $\phi=\{\phi_{1},\phi_{2} \}$ where $\phi_{1} = \{\phi_{11},\phi_{12}\}$ and $\phi_{2} = \{\phi_{21},\phi_{22}\}$. The agents will choose their actions $a_i=[P_{i1c}~P_{i2c}~ \cdots~P_{iQ_{i}c}~\bm{w}_{i1c}~ \bm{w}_{i1p}~\bm{w}_{i2c}~ \bm{w}_{i2p}~\cdots~\bm{w}_{iQ_{i}c}~ \bm{w}_{iQ_{i}p}]$ according to the partial state $O_i$ by following a deterministic policy $a_i = \mu_{\phi_{i}}(O_i)$. To ensure sufficient exploration, we also add a noise vector, whose entries are independent and identically distributed according to $ \mathcal{N}(0,\,\sigma_N^{2}) $ to the deterministic action $a_i = \mu_{\phi_{i}}(O_i)$. Then, the gradient of the expected reward $J(\phi_i)$ for each agent $i$ can be computed as
\begin{align}
   \lefteqn{  \nabla_{\phi_i} J(\phi_i)} & \nonumber\\
  &= \mathbb{E}\left[\nabla_{a_i} \mathcal{Q}_{\theta_i}^{\mu}(s, a_1, a_2)|_{a_i=\mu_{\phi_i}(O_{i})} \nabla_{\phi_i} \mu_{\phi_i}(O_{i})\right],
\end{align}
where $\mathcal{Q}_{\theta_i}^{\mu}(s, a_1, a_2)$ represents the state action value function parameterized by the critic network with $\theta_i$ for each agent. It takes as input the state information $s=(O_1,O_2)$; i.e., channel gains for all users, and the actions $a=(a_1,a_2)$ of all agents, and outputs the $\mathcal{Q}$-value for agent $i$. 

%After indicating states and action, 
Each agent receives a collaborative reward, denoted by $r_{\beta}$, which is a function of the environmental state and actions taken according to state observation. 
\begin{align}\label{eq:reward}r_{\beta} = \beta r_1 + (1-\beta)r_2 , \end{align}
where $ r_1 = R_{1c} + R_{1p}$ and $r_2 = R_{2c} + R_{2p}$ for the case of RSMA, and $\beta$ and $1-\beta$ denote the given weights of the user rates, defined in \eqref{eq:maxrateobj}. MADDPG uses this rate expression to maximize the total discounted return, which is given by
$\mathrm{R} = \sum_{t=0} ^{\infty} \gamma^t r_{\beta,t}^\mu$ where $r_{\beta,t}^\mu$ is the average sum-rate reward obtained under policy $\mu$ at time $t$, and $\gamma \in [0,1]$ denotes the discount factor.

The critic network estimates the $\mathcal{Q}$-value function $\mathcal{Q}_{\theta_i}^{\mu}(s, a_1, a_2)$, which is the expected cumulative reward starting from state $s=(\bm{H}_1, \bm{G}_1, \bm{H}_2, \bm{G}_2)$ and taking a joint action $a$ under policies $\mu=\{\mu_{\phi_{1}},\mu_{\phi_{2}}\}$. MADDPG algorithm employs an experience replay buffer which records the experiences of all agents and stores tuples $<s,a_1,a_2,r_{\beta},s'>$. A mini-batch of $B$ experience tuples ${<s_j, a_{1,j}, a_{2,j}, r_{\beta,j}, s'_j>}_{j=1}^B$ are randomly sampled from the replay buffer $\mathcal{D}$, where $s'_j=(O'_{1,j},O'_{2,j})$ and $r_{\beta,j}$ denote the next state and reward observed after actions $a_{1,j}$, $a_{2,j}$ are taken at state $s_j=(O_{1,j},O_{2,j})$, respectively.

In addition, target networks from the DQN algorithm~\cite{DQN} are adopted to provide stability between actor and critic updates. Target actor networks and critic networks are denoted by $\mu^-$ and $Q^-$ and parameterized by $\phi^-$ and $\theta^-$, respectively.

We update the critic network by minimizing the mean-squared temporal difference (TD) error $\mathcal{L}(\theta_i)$ in sampled mini-batch. 
\begin{equation}
\mathcal{L}(\theta_i) = \frac{1}{B} \sum_{j=1}^B \left(y_j - \mathcal{Q}_{\theta_i}^{\mu}(s_j, a_{1,j} a_{2,j})\right)^2,
\end{equation}
% \begin{equation}
% \mathcal{L}(\theta) = \frac{1}{B} \sum_{j=1}^B \left(y_j - Q^{\mu_i}(s_j, a_j; \theta_i)\right)^2,
% \end{equation}
where the TD target $y_j$ is computed as
\begin{equation}
\begin{split}
y_j &= r_{\beta,j} + \gamma \mathcal{Q}_{\theta_i}^{\mu^{-}}(s'_{j}, 
\mu_{\phi^{-}_{1}}(o'_{1,j}), \mu_{\phi^{-}_{2}}(o'_{2,j})).
\end{split}
\end{equation}
% \begin{equation}
% \begin{split}
% y_j &= r_{\beta,j} + \gamma Q^{\mu_{i-}}(s_{j+1}, 
% \mu_{\phi_{i-}}(s_{1,j+1}), \mu_{\phi_{i-}}(s_{2,j+1}),\\
% &\quad \dots, \mu_{\phi_{i-}}(s_{N,j+1}); \theta_{i-}).
% \end{split}
% \end{equation}
Then, we update the actor network for each agent $i$ by using the deterministic policy gradient as
\begin{equation}
\nabla_{\phi_i} J(\phi_i) \approx \frac{1}{B} \sum_{j=1}^B \nabla_{a_i} \mathcal{Q}_{\theta_i}^{\mu}(s_j, a_{1,j}, a_{2,j}) \nabla_{\phi_i} \mu_{\phi_i}(O_i).\label{nablaa}
\end{equation}
% \begin{equation}
% \nabla_{\phi_i} J(\phi_i) \approx \frac{1}{B} \sum_{i=1}^B \nabla_{a_i} Q^{\mu_i}(s_i, a_i, \dots, a_N; \theta_i) \nabla_{\phi_i} \mu_\phi(s_{i,t}).\label{nablaa}
% \end{equation}
%\todo[inline]{is it $\phi$ or $\phi_i$?}
The target networks are then updated softly to match actor and critic parameters
\begin{align}
\phi^-_{i} &\leftarrow \tau \phi_{i} + (1-\tau) \phi^-_{i}\
&\theta^-_{i} &\leftarrow \tau \theta_i + (1-\tau) \theta^-_{i},
\end{align}
where $0<\tau<1$ is a hyper-parameter controlling the update rate. Finally, we select the best sum-rate over all decoding orders for exhaustive search case. 

The MADDPG with rate-splitting for interference channels is also detailed in Algorithm~\ref{algorithm1}. The hyperparameters of the algorithm are tuned experimentally for the sum-rate maximization problem and are outlined in Table~\ref{tab:mimo_hyper}.

In the next subsection, we enhance the MADDPG algorithm by incorporating decoding order optimization rather than employing exhaustive search. The efficacy of this approach will be demonstrated in Section \ref{sec:sim}.

\subsection{Decoding Order Optimization}
\label{sec:decoding}

In multi-user systems, RSMA utilizes a decoding order through SIC, crucial for determining achievable transmission rates. The decoding order is directly related with the point achieved on the rate region. Therefore, determining the optimal decoding order is also crucial to achieve higher rates. Once the optimal decoding order is determined by the MADDPG framework, this information can be conveyed to the receivers in a few bits, without harming the overall communication rate. %Its dynamic feature enables adaptation to the decoding sequence, ensuring efficient communication amid interference and maintaining a balanced approach to reduce interference without compromising data stream reliability.

%{\color{red}Ensuring receivers decode common messages before extracting private information,}

To estimate the optimal decoding order, which maximizes the sum-rate, an additional actor network is introduced to the multi-agent system.  Let $\eta_i$ denote the decoding order for ${UE}_i$, $i=1,2$. In a scenario with two base stations and two users, where ${UE}_i$ faces two possible choices based on the decoding order, as specified in \eqref{decoding order}. Here, the variable $\eta_i$ takes values of either $0$ or $1$, where $0$ corresponds to option (b), and $1$ corresponds to the decoding order indicated in option (a) in \eqref{decoding order}. 

The new actor network is parameterized by $\phi_0$ and designed to share the same state space and reward function  as the existing ones \eqref{eq:reward}. It observes the state $s=(\bm{H}_1, \bm{G}_1, \bm{H}_2, \bm{G}_2)$ and chooses the decoding order $\eta_i$ for both UEs following the deterministic policy $\mu_{\phi_{0}}(s)$. In this specific scenario, agents extends their decision-making to include additional actions and the new set of joint action becomes $a=[P_{1c}~P_{2c}~ \bm{W}_{1c} ~ \bm{W}_{1p} \bm{W}_{2c} ~ \bm{W}_{2p} ~  \eta_1~ \eta_2]$. Following this decision-making process, each agent is rewarded collaboratively. The collaborative reward is determined by the specific decoding order chosen, the environmental state, and the additional actions taken by the agents.

%However, in this specific scenario, each agent extends its decision-making to include additional actions. The new set of action for agent $i$ is denoted as $a_i=[P_{i1c}~P_{i2c}~ \cdots~P_{iQ_{i}c}~\bm{w}_{i1c}~ \bm{w}_{i1p}~\bm{w}_{i2c}~ \bm{w}_{i2p}~\cdots~\bm{w}_{iQ_{i}c}~ \bm{w}_{iQ_{i}p}~ \eta_i]$, where $\eta_i$ denotes the decoding order. These actions are chosen based on the partial state $O_i$ and follow a deterministic policy, similar to the existing decision-making process.

%The partial state, denoted as $O_i$, plays a crucial role in determining the order of the decoder, resulting in discrete outputs. For example, consider a scenario with two base stations and two users, where ${UE}_i$ faces two possible choices based on the decoding order, as specified in equation \eqref{decoding order}. Here, the variable $\eta_i$ takes values of either 0 or 1, where 0 corresponds to option (b), and 1 corresponds to the decoding order indicated in option (a).

\begin{algorithm}[t]
\caption{MADDPG with RSMA for Weighted Sum-Rate Maximization}
\begin{small}
\begin{algorithmic}
\State Initialize actor networks $\mu_{{\phi_i}}(O_i)$ and critic networks $\mathcal{Q}_{\theta_i}(s_i,a_{i,1},a_{i,2})$ with weights $\theta_i$ and $\phi_i$ %(We will have 2 actors and 2 critics for two base stations.)
\State Initialize target networks $\mu^-$ and $\mathcal{Q}^-$ with weights $\theta^-_{i} \leftarrow \theta_i$ and $\phi^-_{i} \leftarrow \phi_i$
\State Initialize replay buffer $\mathcal{D}$
\For{$episode = 1,\ldots, E$}
%\State Initialize a random process $\mathrm{N}$ for action exploration
\For{$t = 1,\ldots, T$}
\For{each agent $i$}
\State Observe partial state $O_i=(\bm{H}_i, \bm{G}_i)$
\State Select action $a_i=\mu_{\phi_{i}}(O_i)$, 
\State Execute action with exploration noise:
\State $\quad a_i = \mu_{\phi_{i}}(O_i) + \mathcal{N}(0,\,\sigma_N^{2})$.
%\State For each agent $i$, and for each base station $j$ select an action, which will be used to find a precoding vector and a power allocation coefficient for each base station $a_i = \mu_i(s_i|\theta_i) + \pi_t$ according to the policy and exploration noise
\State Observe reward  $r_i$.
\EndFor
\State Observe: 
\State $\quad$ the sum-rate reward $r_{\beta}=\beta r_1+(1-\beta)r_2$, 
\State $\quad$ state
$ s=(O_1~O_2)=(\bm{H}_1, \bm{G}_1, \bm{H}_2, \bm{G}_2)$ and \State $\quad$ next state $s'$
\State Add transition $(s,a_{1},a_{2},r_{\beta},s')$ to $\mathcal{D}$
\State Sample a minibatch of $B$ transitions:
\State $\quad (s_j,a_{1,j},a_{2,j},r_{\beta,j},s_j')$ from $\mathcal{D}$
\State Compute target action:
\State $\quad a'_j = \mu^-(o'_j)$
\State Compute target $\mathcal{Q}$-value:
\State $\quad y_j = r_{\beta,j} + \gamma \mathcal{Q}^-(s'_j,a'_{1,j},a'_{2,j})$
\State Update each critic by minimizing the loss: 
\State $\quad L(\theta_i) = \frac{1}{B} \sum_{j=1}^B(y_j - \mathcal{Q}_{\theta_i}^\mu(s_j,a_{i,1},a_{i,2}))^2$
\State Update each actor using the sampled policy gradient:
\State $\quad \nabla_{\phi_i} J(\phi_i) \approx \frac{1}{B} \sum_{j=1}^B \nabla_{a_i} \mathcal{Q}_{\theta_i}^{\mu}(s_j, a_{1,j}, a_{2,j}) 
\nabla_{\phi_i} \mu_{\phi_i}(O_i)$
% \parbox[t]{\dimexpr\linewidth-\algorithmicindent}{%
% $\nabla_{\phi_i} J(\phi_i) \approx \frac{1}{B} \sum_{j=1}^B \nabla_{a_i} Q_{\theta_i}^{\mu}(s_j, a_{1,j}, a_{2,j}) 
% \nabla_{\phi_i} \mu_{\phi_i}(o_i)$
% }
\State Soft update target networks:
%$\theta'_i \leftarrow \tau\theta_i + (1-\tau)\theta'_i$, $\phi^-_{i} \leftarrow \tau \phi_{i} + (1-\tau) \phi^-_{i}$
%\vspace{-0.2cm}
% \begin{align}\theta^-_{i} &\leftarrow \tau\theta_i + (1-\tau)\theta^-_{i} \nonumber \\
%  \phi^-_{i} &\leftarrow \tau \phi_{i} + (1-\tau) \phi^-_{i} \nonumber \end{align}
\State $\quad ~~ \theta^-_{i} \leftarrow \tau\theta_i + (1-\tau)\theta^-_{i}, ~~
 \phi^-_{i} \leftarrow \tau \phi_{i} + (1-\tau) \phi^-_{i}$
 % $$\theta^-_{i} \leftarrow \tau\theta_i + (1-\tau)\theta^-_{i} $$
% $$ \phi^-_{i} \leftarrow \tau \phi_{i} + (1-\tau) \phi^-_{i}$$
\EndFor
\EndFor
\end{algorithmic}
\end{small}
\label{algorithm1}
\end{algorithm}

\subsection{Imperfect Channel State Information}

In this subsection, channel estimation errors, a practical consideration in real-world scenarios, is systematically explored. The examination of these errors encompasses two distinct modalities: the first, wherein the estimation error remains the same irrespective of the SNR value, and the second, where the estimation error decreases with increasing SNR. The estimated channels can be represented as
\begin{equation} \label{channel_estimation}
 \tilde{\bm{H}}_i = \bm{H}_i + \bm{E}_i, \quad \tilde{\bm{G}}_j = \bm{G}_j + \hat{\bm{E}}_j
\end{equation}
where the resulting estimation error terms, denoted by $\bm{E}_i$ and $\hat{\bm{E}}_j$, arise from the disparity between the estimated channel coefficients, represented by $\tilde{\bm{H}}_i$ and $\tilde{\bm{G}}_j$, and the actual channel coefficients denoted as $\bm{H}_i$ and $\tilde{\bm{G}}_j$. It is important to note that the matrices $\bm{E}_i$, $\bm{H}_i$, and $\tilde{\bm{H}}_i$ all belong to the complex field and have dimensions $N_{i} \times M_i$; i.e. $ \mathbb{C}^{N_{i} \times M_i}$. For the second case, where estimation errors decrease with increasing SNR, each entry $[\bm{E}_i]{(x,y)}$, $x=1,\ldots, N_i$, $y = 1,\ldots, M_i$, is determined as $\frac{SNR^{-0.6}}{5}\mathcal{CN}(0, \sigma^2)$, where $\sigma^2 = 1$. Similarly, the matrices $\hat{\bm{E}}_j$, $\bm{G}_j$, and $\tilde{\bm{G}}_j$ are all in $\mathbb{C}^{N_i\times M_j}$, $i\neq j$. Each entry in $\hat{\bm{E}}_i$ is drawn from $\frac{SNR^{-0.6}}{5}\mathcal{CN}(0, \sigma^2)$ with $\sigma^2=1$. In the case of fixed channel estimation errors, each entry in $\bm{E}_i$ or $\hat{\bm{E}}_j$
will be distributed according to $\frac{10^{-0.6}}{5}\mathcal{CN}(0, \sigma^2)$ with $\sigma^2=1$ in Section~\ref{sec:sim}.

It is crucial to highlight that the estimation error solely influences the computation of precoders, while the rates are computed employing the exact channel coefficients. The actor network in each agent incorporates estimated channel coefficients, whereas the reward function is formulated based on the exact channel coefficients.

%\todo[inline]{Nuri, Table 1'i eski makalelerden kopyaladim. Orijinali DQN icin, bunu bizim algoritmammiza ve kodumuzdaki parametrelere uyarlar misin? Ya tabloya ya da text'e de yine suna benzer cumleler yazmak gerekiyor: "We construct a feed-forward neural network of ... layers,  one of which is the hidden layer with ... neurons.  "}

\begin{table}[t]
\caption{Hyperparameters of MADDPG algorithm for MIMO}\label{tab:mimo_hyper}
\begin{footnotesize}
\begin{center}
 \label{table_MADDPG3}
\begin{tabular}{ |c|c|c|c| } 
\hline
 Parameter & Value & Parameter & Value \\ 
 \hline \hline
 discount factor $\gamma$ & 0.99 & optimizer & Adam \\ 
 \hline
 minibatch size $B$ & 128 & loss function & MSE loss\\ 
 \hline
 replay memory length $D$  & 15000 & no. of connected layers & 5 \\ \hline 
 activation function & ReLU & learning rate   & $5 \times 10^{-5}$ \\ \hline 
   hidden size & 64  & episode length $T$ & 200  \\ 
\hline update rate $\tau$ & 0.01 & exploration noise $\sigma_N^{2}$ & 0.1 \\
\hline number of episodes $E$ & 12000 & weight of user rates $\beta$ & 0.5  \\
\hline   
   % episode length $T$ & 250  & $\epsilon_{min}$ & 0.01 \\ \hline 
\end{tabular}
\end{center}
\end{footnotesize}
\end{table}

\subsection{MADDPG with no Rate-Splitting}
\label{sec:DRL_withoutRSMA}
If there is no rate-splitting, our scheme reduces to the one in \cite{lee2021multi}. In this case, there is no common message, $\bm{b}_{ic}=\emptyset$, and $\bm{W}_{ic}$ is an all zero matrix. The one in that for MADDPG without rate-splitting,  \eqref{eq:reward} can be modified by using $r_1 = R_{1}$ and $r_2 = R_{2}$ that are given in \eqref{eq:rate1_no_rs} and \eqref{eq:rate2_no_rs}. Also, since we only optimize precoders, but not $\bm{P}_{ikc}$, we use actions only for precoder evaluation. Then, the rates achieved by this scheme for the MIMO case become
% \begin{align}\label{eq:nineteen}R_{1} = \log\left(1 + \dfrac{|\bm{h}_{1}\bm{w}_{1}^{RL}|^{2}} {{|\bm{{g}_{2}}\bm{w}_{2}^{RL}|^{2}} + N_{0}}\right)  \end{align}
% \begin{align}\label{eq:ninetee1n}R_{2} = \log\left(1 + \dfrac{|\bm{h}_{2}\bm{w}_{2}^{RL}|^{2}} {{|\bm{g}_{1}\bm{w}_{1}^{RL}|^{2}} + N_{0}}\right).  \end{align}
\begin{eqnarray} \nonumber
    R_{1}^{\pi} &=& \log_2\det\Bigl(\bm{I}_{N_1} + {(\bm{H}_{1}\bm{W}_{1}^{\pi}\bm{W}_{1}^{\pi,H}\bm{H}_{1}^{H})  }\\ && \qquad  \quad {({{\bm{{G}}_{2}\bm{W}_{2}^{\pi}\bm{W}_{2}^{\pi,H}\bm{{G}}_{2}^{H}} + N_{0}\bm{I}_{N_1}})}^{-1}\Bigl) \label{eq:rate1_no_rs} \\ \nonumber
    R_{2}^{\pi} &= &\log_2\det\Bigl(\bm{I}_{N_2} + {(\bm{H}_{2}\bm{W}_{2}^{\pi}\bm{W}_{2}^{\pi,H}\bm{H}_{2}^{H})} \\ && \qquad  \quad {({{\bm{{G}}_{1}\bm{W}_{1}^{\pi}\bm{W}_{1}^{\pi,H}\bm{{G}}_{1}^{H}} + N_{0}\bm{I}_{N_2}})}^{-1}\Bigr)\label{eq:rate2_no_rs} 
\end{eqnarray}
% \begin{equation}
% \begin{aligned}\label{eq:rate1_no_rs}
% R_{1}^{\pi} &= \log_2\det\Bigl(\bm{I}_{N_1} + {(\bm{H}_{1}\bm{W}_{1}^{\pi}\bm{W}_{1}^{\pi,H}\bm{H}_{1}^{H})  }\\ & \qquad \quad \quad \quad {({{\bm{{G}}_{2}\bm{W}_{2}^{\pi}\bm{W}_{2}^{\pi,H}\bm{{G}}_{2}^{H}} + N_{0}\bm{I}_{N_1}})}^{-1}\Bigl) 
% \end{aligned}
% \end{equation}
% \begin{equation}
% \begin{aligned}\label{eq:rate2_no_rs}
% R_{2}^{\pi} &= \log_2\det\Bigl(\bm{I}_{N_2} + {(\bm{H}_{2}\bm{W}_{2}^{\pi}\bm{W}_{2}^{\pi,H}\bm{H}_{2}^{H})} \\ & \qquad \quad \quad \quad {({{\bm{{G}}_{1}\bm{W}_{1}^{\pi}\bm{W}_{1}^{\pi,H}\bm{{G}}_{1}^{H}} + N_{0}\bm{I}_{N_2}})}^{-1}\Bigr) 
% \end{aligned}
% \end{equation}
where $\pi=\{drl\}$ indicates the precoders for MADDPG without rate-splitting. 

\subsection{Complexity Analysis of the Proposed Algorithm}
\label{sec:complexity}

Training in MADDPG is the phase where agents interact with their environment, improve their decision-making policies, and adjust the parameters of the actor and critic networks. This phase is the most computationally expensive because it requires a lot of calculations to learn the best policies and value functions. Key steps in training involve several important processes. First, each agent stores its experiences in a replay buffer, which includes states, actions, and rewards from previous time steps. Then, each agent updates its policy by computing gradients based on feedback from the critic network, which estimates the value of state-action pairs. The critic network itself updates its value estimates using a loss function that incorporates the Bellman equation. Finally, agents must balance exploration and exploitation, requiring multiple interactions with the environment to explore new strategies while exploiting known ones. Specifically, the training complexity can be expressed as
% \begin{align}\label{eq:complexity_exp} 
% O\Bigg( &E \cdot \Big( N_{\text{agents}} \cdot K \cdot Q_a^2 + L^2 \cdot M \cdot K \cdot B^2 \Big) \nonumber \\
% &+ T \cdot D + N_{\text{agents}} \cdot \Big( K \cdot Q_a^2 + L^2 \cdot M \cdot K \cdot B^2 \Big) \Bigg)
% \end{align} 
\begin{eqnarray}  
\lefteqn{O \left( E \left( N_{\text{agents}}  K  Q_a^2 + L^2  M  K  B^2 \right) 
+ T  D \right.  } \notag \\
&\quad \quad \quad \quad \quad \quad \quad+ N_{\text{agents}} \left. \left( K  Q_a^2 + L^2  M  K  B^2 \right) \right)\label{eq:complexity_exp}
\end{eqnarray} where
% \begin{itemize}
%   \item \( E \) is the number of training episodes or time steps.
%   \item \( N_{\text{agents}} \) is the number of agents.
%   \item \( K \) is the action space size.
%   \item \( Q_a \) is the size of the Q-value function.
%   \item \( L \) is the number of layers in the network.
%   \item \( M \) is the number of units(hidden size) in each layer.
%   \item \( B \) is the batch size.
%   \item \( T \) is the episode length.
%   \item \( D \) is the replay memory length.
% \end{itemize}
\( E \) is the number of training episodes or time steps,  \( N_{\text{agents}} \) is the number of agents, \( K \) is the action space size, \( Q_a \) is the size of the $\mathcal{Q}$-value function, \( L \) is the number of layers in the network, \( M \) is the number of units(hidden size) in each layer, \( B \) is the batch size, \( T \) is the episode length, and \( D \) is the replay memory length.

The expression \eqref{eq:complexity_exp} involves several key components. The forward pass complexity is \(O(N_{\text{agents}}  K  Q_a^2)\) for each agent, which is required to calculate $\mathcal{Q}$-values and process them through the network. The network layers' complexity is \(O(L^2  M  K  B^2)\), accounting for the complexities of the neural network layers in both the actor and the critic networks. The experience replay complexity is \(O(T D)\), representing the effort required to store and sample experiences from the replay buffer during training. The critic update complexity is \(O(N_{\text{agents}}  (K  Q_a^2 + L^2  M  K  B^2))\), which includes the complexity of calculating the Bellman equation and performing back-propagation for the critic network. Together, these components provide a comprehensive picture of the computational complexity of MADDPG, covering the training process, experience replay, and critic updates, all scaled by the number of agents and training steps. The forward calculation complexity is low because, once the model is trained, the forward pass becomes computationally efficient. It simply involves a feed forward operation through the actor and critic networks to select actions and estimate values. Since there are no gradients to compute or weights to update during the forward pass, the process is much simpler. Essentially, the forward pass only requires matrix multiplications for each network, which is far less computationally demanding compared to the training phase. This is a key reason why execution complexity is low across most deep learning algorithms once the model has been trained. 

In MADDPG, after training, the model is deployed in a distributed setting where each agent executes its own policy in parallel, independently interacting with the environment and selecting actions based on the trained model. The key steps in distributed execution include parallel execution, where multiple agents execute their policies concurrently, each having its own local environment; action selection, where each agent uses the trained actor model to predict actions based on its local state; environment interaction, where agents interact with the environment and obtain feedback (reward and next state); and state update, where each agent updates its local state in real time based on the environment's response.

The complexity of distributed execution can be expressed as
\begin{align}\label{eq:execution_complexity} 
O(N_{\text{agents}}  (Q_a + Q_c))
\end{align}
where \(N_{\text{agents}}\) is the number of agents in the distributed system, and \(Q_a\) and \(Q_c\) are the output sizes of the actor and critic networks for action and value predictions. 
The distributed execution complexity is low due to several reasons. First, parallelization allows each agent to independently perform a forward pass to select its action, with no interactions between agents during this phase, except for sharing the environment. Second, no training is involved in this phase, as there is no back propagation or gradient update; agents only compute actions based on the trained model. Lastly, real-time execution makes the complexity low per agent, and the load can be efficiently scaled by distributing tasks across agents.

Complexity of simple precoders like ZF is primarily determined by the need to solve matrix inversion problems, which has a computational cost of \( O(N_i^3 + N_i^2 K) \), where \( N_i \) is the number of antennas and \( K \) is the number of users. Advanced RSMA schemes, which require convex optimization, scale to \( O(N_i^3 K^3) \). However, this does not account for the added complexity of decoding order selection, which can be substantial, especially when exhaustive search methods are used. This can add \( O(K!) \) to the overall complexity, though practical implementations often employ heuristic methods that reduce this to \( O(K^2) \). When compared with MADDPG, which operates in a distributed manner, the complexity is quite different. MADDPG's execution complexity scales to \eqref{eq:execution_complexity}  represent the complexities of each agent's action and critic evaluation, respectively. This distributed execution of MADDPG results in a significantly lower overall complexity compared to traditional RSMA optimization, such as weighted MMSE \cite{yalcin2018downlink}, especially as the number of users or agents increases, demonstrating the efficiency of MADDPG in multi-agent settings relative to the centralized nature of RSMA.

%To reduce the training time of deep learning models like MADDPG, specialized hardware accelerators such as Xilinx AI Cores \cite{xilinx_ai_cores}, Google’s Tensor Processing Unit (TPU) \cite{google_tpu}, NVIDIA’s Tensor Cores \cite{nvidia_tensors}, and Intel’s Nervana NNP \cite{intel_nervana} are highly effective. These devices are specifically designed to accelerate the computationally intensive operations involved in model training, including backpropagation, gradient calculations, and matrix multiplications. TPUs and Tensor Cores are optimized for tensor operations, speeding up the calculations involved in training deep neural networks by performing large-scale matrix multiplications and convolutions much faster than traditional CPUs. Additionally, Intel's Nervana NNP and Xilinx Versal AI Core are engineered for AI workloads, providing high throughput and parallel processing capabilities that enhance the efficiency of training deep learning models. By offloading complex computations to these specialized devices, training time can be significantly reduced, enabling faster model convergence and more efficient deployment.

\section{Benchmark Precoding Schemes}\label{sec:bench}

In this section, we will be explaining the benchmark precoding schemes maximum ratio transmission (MRT) \cite{jorswieck2008complete}, zero-forcing (ZF) precoding \cite{jorswieck2008complete}, and leakage-based precoding \cite{sadek2007leakage}. We will also compare with the upper bounds \cite{etkin2008gaussian, sato1981capacity, karmakar2013capacity} on interference channels.

\subsection{Maximum Ratio Transmission}

MRT is employed at the transmitter side, where transmit antenna weights are matched to the channel \cite{jorswieck2008complete},\cite{goldsmith2005wireless}. This way, the maximum received SNR is attained at the intended receivers. This process takes advantage of the spatial diversity offered by multiple antennas at both the transmitter and receiver ends. By adjusting the transmission weights in this manner, MRT aims to maximize the received signal power, effectively exploiting the available spatial dimensions and enhancing the system's overall performance in terms of reliability and data throughput. However, MRT does not take interference into consideration. In MRT there is no rate-splitting and there is no common message, $\bm{b}_{ic}=\emptyset$, $\bm{W}_{ic}$ is an all zero matrix, $\bm{W}_i^{mrt}=\bm{W}_{ip}$ and for the MIMO case the precoder expression is given as
\begin{align}\label{eq:mrt_mimo} 
{\bm{W}_i^{mrt}} = {\bm{H}_i^{H}}.  
\end{align}

\subsection{Zero-Forcing}
As in MRT, there is no rate-splitting in ZF transmission, and the transmitters aim to eliminate interferences among data streams by setting the transmission weights such that there is no interference caused on the receiving antennas. This is achieved by using the pseudo-inverse of the channel matrix to create a null space for the interference; i.e., by projecting input data symbols on the null space of $ \bm{G_{i}}$. By nullifying the interference, ZF seeks to improve the reliability of data transmission without causing mutual interference among the multiple antennas, thereby enhancing the overall system performance in terms of throughput and signal quality. However, ZF might be sensitive to noise and can lead to amplification of noise in the process of eliminating interference. As a result, for ZF $\bm{b}_{ic}=\emptyset$, $\bm{W}_{ic}$ is an all zero matrix, and $\bm{W}_i^{zf}=\bm{W}_{ip}$, where
\begin{align}\label{eq:zf_mimo}{\bm{W}_{i}^{zf}} = {({\bm{G}_{i}^{H}}{\bm{G}_{i}})^{-1}}{\bm{H}_{i}^{H}}.  
\end{align}

\subsection{Leakage-based Precoding}

SLNR precoding technique is designed to optimize signal transmission in multi-user communication systems by minimizing interference among users while considering the system's noise. Unlike traditional SNR-based approaches, SLNR focuses on minimizing both interference and noise to enhance the overall signal quality. In SLNR precoding, the precoding matrix is computed to maximize the desired signal power while minimizing the interference caused to other users. It aims to maintain a high signal-to-leakage-plus-noise ratio for the intended receiver, hence reducing interference while considering the system noise level. This precoding strategy is particularly useful in multi-user scenarios where reducing interference among users is crucial to improve overall system performance. In other words, leakage is a measure of how
much signal power leaks into the other users. In this precoding scheme, the aim is to maximize the SLNR \cite{sadek2007leakage}. 

For a MIMO system the $k$th column of ${\bm{W}_{i}^{slnr}}$, ${\bm{w}_{ik}^{slnr}}$, is equal to the eigenvector that corresponds to the largest eigenvalue of 
 ${\left({(N_{0}\bm{I} +{\bm{G}_{ik}^{H}\bm{G}_{ik}})^{-1}\bm{H}_{ik}^{H}\bm{H}_{ik}}\right)}$ where $\bm{G}_{ik}$ and $\bm{H}_{ik}$ represent the $k$th column of $\bm{G}_{i}$ and $\bm{H}_{i}$ respectively.

\subsection{Interference Channel Upper Bounds}\label{subsec:upper}

In \cite{sato1981capacity} and \cite{etkin2008gaussian}, the authors suggest upper bounds for single antenna interference channels. For the single antenna case, we will use the upper bound given in \cite[(58)-(71)]{etkin2008gaussian} for weak and mixed interference conditions. In this context, for $i=1,2$, let $\bm{H}_i=h_i$  and $\bm{G}_j= g_j$. Then, the signal-to-noise ratio (SNR) of user $i$, $\mathrm{SNR}_i$, and interference-to-noise ratio (INR) for ${UE}_i$ caused by $UE_j$, $\mathrm{INR}_i$, of \cite[(58)-(71)]{etkin2008gaussian} respectively become $\mathrm{SNR}_i=\left|h_{i}\right|^2 P_i / N_0$ and $\mathrm{INR}_i=\left|g_{j}\right|^2 P_j / N_0$. Furthermore, the weak interference condition is defined as $\mathrm{INR}_1 < \mathrm{SNR}_2$ and $\mathrm{INR}_2 < \mathrm{SNR}_1$, while the mixed interference condition encompasses scenarios where either $\mathrm{INR}_1 \geq \mathrm{SNR}_2$ and $\mathrm{INR}_2 < \mathrm{SNR}_1$ or $\mathrm{INR}_1 < \mathrm{SNR}_2$ and $\mathrm{INR}_2 \geq \mathrm{SNR}_1$.  %Here, $P_1$ and $P_2$ represent the power levels of the first and second users, respectively. Bu tanimli zaten.
Finally, the strong interference condition occurs, when $\mathrm{INR}_1 > \mathrm{SNR}_2$ and $\mathrm{INR}_2 > \mathrm{SNR}_1$. The interference channel becomes equivalent to a compound multiple access channel, in which both users decode all messages. In this case the symmetric rate point is given in \cite[(33)]{etkin2008gaussian} and the whole region is expressed in \cite[(5)]{sato1981capacity}.

%For the scenario with strong interference, we employ the upper bound evaluation from the reference \cite[(5)]{sato1981capacity}. In this paper, the variables $\alpha$ and $\beta$ denote the degree of interference present in the first and second output signals, namely $y_1$ and $y_2$. However, as we lack specific scaling parameters, we define the strong interference condition in our case as $\mathrm{INR}_1 > \mathrm{SNR}_2$ and $\mathrm{INR}_2 > \mathrm{SNR}_1$. We handle our computations by assuming $\alpha = \beta = 1$ within this framework. Additionally, we adopt the simplification where $N_1 = N_2 = N_0$ for this analysis.

With these definitions, for each channel realization, we first determine whether the channel is a weak, mixed or strong interference channel and then directly apply the corresponding bounding equations in \cite[(58)-(71)]{etkin2008gaussian} or \cite[(5)]{sato1981capacity} for the SISO case. 
 
 %When we are averaging over different channel conditions in the next section, for each channel realization we check the interference condition (weak, mixed or strong), apply the appropriate bound and then take the average. 

When some of the devices have multiple antennas, the above bounds are not directly applicable and we adapt the upper bound calculated in \cite{karmakar2013capacity}. For a comprehensive presentation, we present the bound as follows: 
\begin{align} \nonumber
R_1 & \leq \log_2 \operatorname{det}\left[\bm{I}_{N_1}+\rho_{11} \bm{H_{1}} \bm{H_{1}}^{H}\right] \\ \nonumber
R_2 & \leq \log_2 \operatorname{det}\left[\bm{I}_{N_2}+\rho_{22} \bm{H_{2}} \bm{H_{2}}^{H}\right] \\ \nonumber  
R_1+R_2 & \leq \log_2 \operatorname{det}\left[\bm{I}_{N_2}+  \rho_{12} \bm{G_{1}} \bm{G_{1}}^{H}+ \rho_{22} \bm{H_{2}} \bm{H_{2}}^{H}\right] \\ \nonumber &  + \log_2 \operatorname{det}\left[\bm{I}_{N_1}+\rho_{11} \bm{H_{1}} \bm{K_{1}} \bm{H_{1}}^{H}\right] \\ \nonumber
R_1+R_2 & \leq \log_2 \operatorname{det}\left[\bm{I}_{N_1}+\rho_{21} \bm{G_{2}} \bm{G_{2}}^{H}+\rho_{11} \bm{H_{1}} \bm{H_{1}}^{H}\right] \\ \nonumber &  +  \log_2 \operatorname{det}\left[\bm{I}_{N_2}+\rho_{22} \bm{H_{2}} \bm{K_{2}} \bm{H_{2}}^{H}\right] \\ \nonumber 
R_1+R_2 & \leq \log_2 \operatorname{det}\left[\bm{I}_{N_1}+\rho_{21} \bm{G_{2}} \bm{G_{2}}^{H}+\rho_{11} \bm{H_{1}} \bm{K_{1}} \bm{H_{1}}^{H}\right] \\ \nonumber & + \log_2 \operatorname{det}\left[\bm{I}_{N_2}+\rho_{12} \bm{G_{1}} \bm{G_{1}}^{H}+\rho_{22} \bm{H_{2}} K_2 \bm{H_{2}}^{H}\right] \\ \nonumber
2 R_1+R_2 & \leq \log_2 \operatorname{det}\left[\bm{I}_{N_1}+\rho_{21} \bm{G_{2}} \bm{G_{2}}^{H}+\rho_{11} \bm{H_{1}} \bm{H_{1}}^{H}\right]  \\ \nonumber &  + \log_2 \operatorname{det}\left[\bm{I}_{N_1}+\rho_{11} \bm{H_{1}} \bm{K_{1}} \bm{H_{1}}^{H}\right] \\ \nonumber & + \log_2 \operatorname{det}\left[\bm{I}_{N_2}+\rho_{12} \bm{G_{1}} \bm{G_{1}}^{H}+\rho_{22} \bm{H_{2}} \bm{K_{2}} \bm{H_{2}}^{H}\right] \\ \nonumber
R_1+2 R_2 & \leq \log_2 \operatorname{det}\left[\bm{I}_{N_2}+\rho_{12} \bm{G_{1}} \bm{G_{1}}^{H}+\rho_{22} \bm{H_{2}} \bm{H_{2}}^{H}\right]  \\ \nonumber &  + \log_2 \operatorname{det}\left[\bm{I}_{N_2}+\rho_{22} \bm{H_{2}} \bm{K_{2}} \bm{H_{2}}^{H}\right] \\ \label{mimo_any_1} & + \log_2 \operatorname{det}\left[\bm{I}_{N_1}+\rho_{21} \bm{G_{2}} \bm{G_{2}}^{H}+\rho_{11} \bm{H_{1}} \bm{K_{1}} \bm{H_{1}}^{H}\right]
\end{align}
where
\begin{eqnarray}
\bm{K}_i  & = & \left(\bm{I}_{M_i}+\rho_{i j} \bm{G_{i}}^{H} \bm{G_{i}}\right)^{-1}, \quad 1 \leq i \neq j \leq 2 \\
\rho_{i i} & =& \frac{ \operatorname{Tr}\left(\bm{H_{i}} \bm{\mathbb{Q}_{i}} \bm{H_{i}}^{H}\right)}{N_i}, \quad i = 1,2, \\
\rho_{ij} & =& \frac{ \operatorname{Tr}\left(\bm{G_{i}} \bm{\mathbb{Q}_{i}} \bm{G_{i}}^{H}\right)}{N_i}, \quad 1 \leq i \neq j \leq 2.
\end{eqnarray}
Finally, the covariance matrix is represented as $\bm{\mathbb{Q}_{i}} = \mathbb{E}\left(\bm{x_{i }} \bm{x_{i}}^{H}\right)$.

\subsection{No Interference}

For this case, interference terms are assumed to be 0 to obtain a trivial upper-bound. The rates achieved by each user for a MIMO system is then written as 
\begin{align}\label{eq:noint1_mimo}R_{1} = \log\det\left(\bm{I}_{N_1} + {(\bm{H}_{1}\bm{W}_{1}\bm{W}_{1}^{H}\bm{H}_{1}^{H})N_{0}^{-1}}  \right),  \end{align}
\begin{align}\label{eq:noint2_miso}R_{2} = \log\det\left(\bm{I}_{N_2} + (\bm{H}_{2}\bm{W}_{2}\bm{W}_{2}^{H}\bm{H}_{2}^{H})N_0^{-1}  \right).  \end{align}

\section{Simulation Results}
\label{sec:sim}

In this section, we delve into an exhaustive examination of simulation results for MADDPG with RSMA. This method is meticulously compared against its subset, the MADDPG with no rate-splitting \cite{lee2021multi}, offering valuable insights into the role of rate-splitting in shaping the overall performance. Moreover, we extend our investigation to encompass a comprehensive set of benchmark schemes outlined in Section~\ref{sec:bench}, covering a spectrum of scenarios including  SISO, MISO and MIMO configurations. This extensive analysis allows us to examine the proposed scheme's adaptability and efficacy across diverse communication scenarios.

%The benchmark schemes considered in our simulations span well-established methodologies for interference channels with multiple antennas. The comparative study systematically evaluates the proposed MADDPG with rate-splitting against these benchmarks, unraveling the nuanced intricacies of each scheme in  SISO, MISO and MIMO cases. By conducting such a comprehensive set of simulations, our aim is to provide a holistic understanding of the strengths, weaknesses, and applicability of each scheme across a range of communication scenarios. This thorough examination serves as a critical step in elucidating the effectiveness and robustness of the proposed MADDPG with rate-splitting, showcasing its relative performance against alternative methodologies in diverse and challenging communication environments.

Our primary objective is to maximize the weighted sum-rate in a two-user interference channel by optimizing both the power allocation coefficients and the precoders. Maximizing the sum-rate in RSMA is crucial as it directly correlates with the overall efficiency and throughput of the communication system. A higher sum-rate implies the ability to transmit more information per unit time. %, leading to improved data transfer capabilities and enhanced network performance. 
The MADDPG algorithm provides a framework for the coordinated learning of multiple agents to optimize the sum-rate. MADDPG enables these agents to adapt their strategies collaboratively, ensuring a balanced and effective allocation of resources, such as power and decoding orders, to maximize the sum-rate. This optimization process, particularly in the context of RSMA, introduces inherent complexities. The nature of RSMA, with its simultaneous consideration of common and private streams, renders traditional optimization approaches less effective, making the utilization of reinforcement learning approaches appealing. In addition to maximizing the sum-rate, our study will delve into the performance implications of decoding order estimation. Furthermore, we will rigorously examine the system's resilience in the face of channel estimation errors, a crucial consideration in practical communication scenarios. This multifaceted exploration aims to provide a thorough understanding of the proposed model's capabilities, particularly in the complex landscape of RSMA.

The instantiation of the MADDPG algorithm was conducted within the PyTorch 1.9.1 framework. The training architecture incorporates four fully connected layers for both the critic and the actors, providing a robust and sophisticated foundation for the learning process. To assess the algorithm's convergence, extensive monitoring was conducted over 12,000 episodes for MIMO, 4,000 episodes for MISO, and 2,400 episodes for SISO cases, with each episode consisting of 200 time steps. This comprehensive evaluation resulted in the acquisition of rates across varying SNR scenarios. Within the simulated environment, the channel coefficients $\bm{H}_i$, and $\bm{G}_i$ for $i=1,2$  are assumed to follow an independent and identically distributed circularly symmetric complex Gaussian distribution with zero mean and unit variance. Irrespective of the antenna configuration, be it SISO, MISO, or MIMO, the total power $P_i$ in (\ref{eq:Pisum}) is consistently set to 1.  %This modeling paradigm aligns with established conventions in the representation of wireless communication channels. 

Our investigation will encompass a comprehensive analysis of our communication scheme across diverse SNR regimes, specifically considering the case where SNR is defined as the reciprocal of the noise power $N_0$ (i.e., SNR = 1/$N_0$), given that the signal power $P_i$ is fixed at 1. %This SNR parameterization is consistent with the power constraint considerations outlined in Section~\ref{sec:system_model}, where $N_0$ represents the noise power. 
To maintain uniformity and adhere to power constraints, the precoders employed in each scheme will undergo normalization by their respective magnitudes. This normalization process ensures that the power constraints are consistently satisfied across various SNR scenarios, facilitating a systematic evaluation of the scheme's performance under different SNR conditions.

\begin{figure}[t]
    \centering
    \includegraphics[width=90mm]{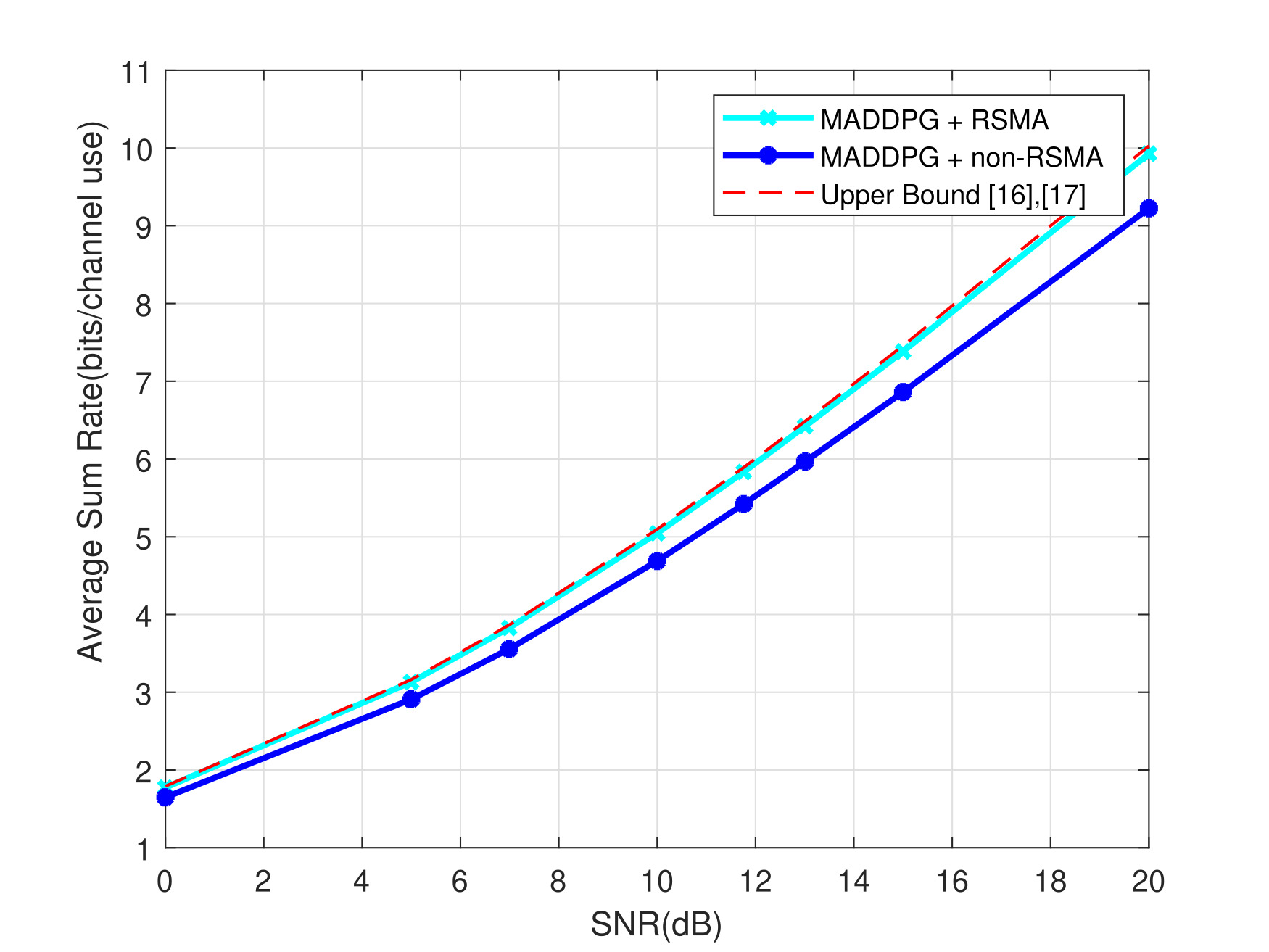}
    \caption{Average sum-rate achieved by MADDPG and the upper bound due to \cite{sato1981capacity}  and \cite{etkin2008gaussian} for $M=1$ and $Q=1$. The MADDPG curves are obtained by averaging 25 runs, each having 200 time steps after the algorithm achieves convergence.}
    \label{Figure:single_antenna}
\end{figure}

Fig. \ref{Figure:single_antenna} illustrates the average sum-rate as a function of SNR for the two-user interference channel for the SISO case; i.e. $M_i=1$, $N_i=1$, $i=1,2$. %In this setup, the system formulation is based on the configuration described in Section \ref{sec:system_model}, while the solution methodology aligns with the structure outlined in Section \ref{sec:MADDPG}. 
The objective is to discern the impact of rate-splitting on performance. The plot includes curves for the MADDPG algorithm both with and without rate-splitting, alongside the upper bound elucidated in Section~\ref{sec:bench}. The MADDPG + RSMA curves are derived from averaging 25 runs, each consisting of 200 time steps, post-convergence of the algorithm. Notably, the results indicate that MADDPG with rate-splitting can attain the upper bound, signifying its efficacy in approaching theoretical limits. Furthermore, as SNR increases, the disparity between the performances of MADDPG with and without rate-splitting becomes more pronounced. This observation underscores the significance of rate-splitting as a strategic mechanism for augmenting the sum-rate, particularly in scenarios characterized by high SNR. 

\begin{figure}[t]
    \centering
    \includegraphics[width=90mm]{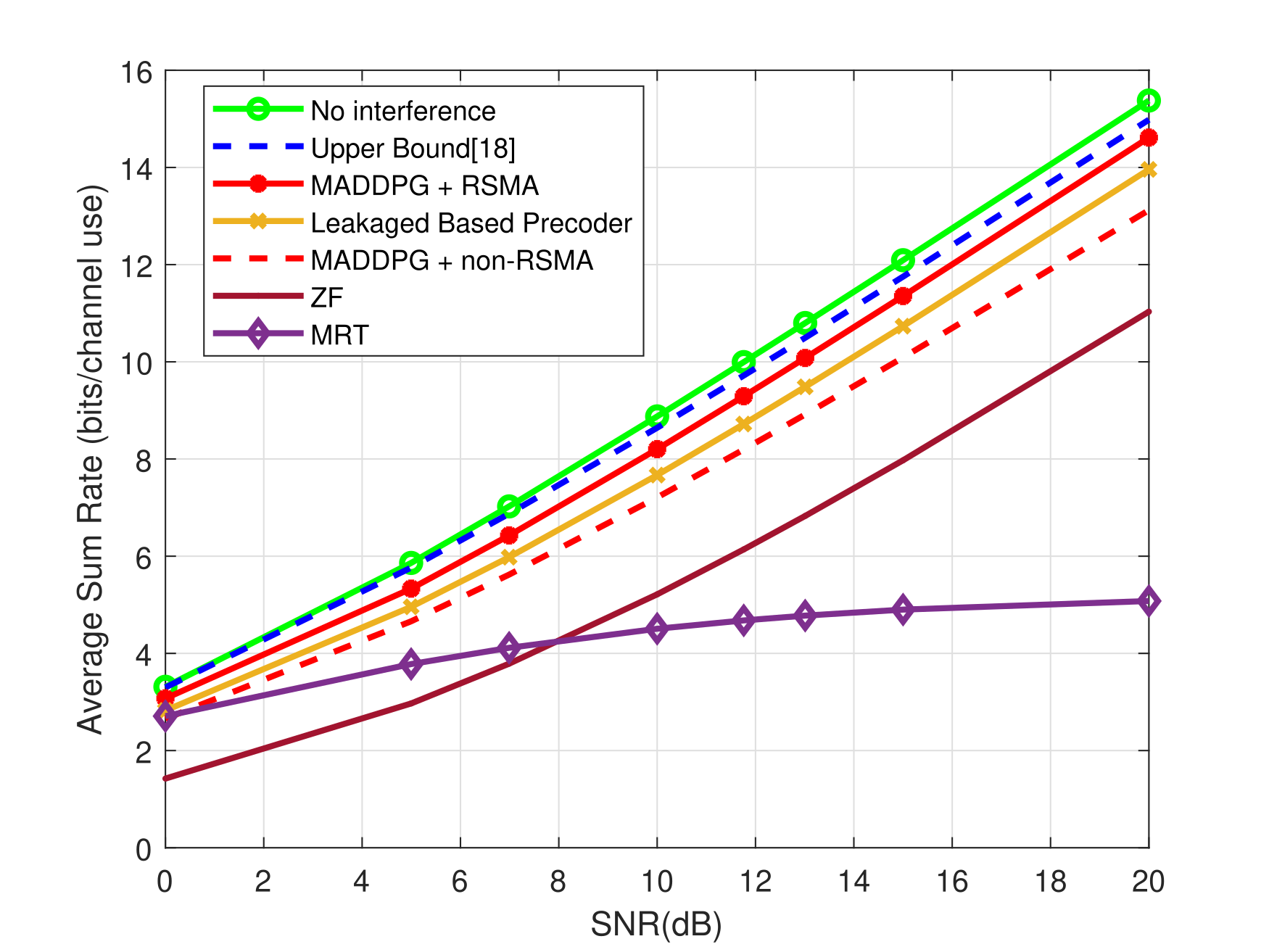}
    \caption{Average sum-rate achieved by MADDPG and the benchmark schemes for $M=3$ and $Q=1$. The MADDPG curves are obtained by averaging 50 runs, each having 1000 time steps after the algorithm achieves convergence.}
    \label{Figure:multi_antenna}
\end{figure}

In Fig.~\ref{Figure:multi_antenna}, we present a detailed analysis of the average sum-rate results for $M = 3$ and $Q=1$. The comparison involves multiple schemes, including MADDPG with and without rate-splitting, MRT, ZF, and leakage-based precoding. Additionally, the upper bound, as defined in \cite{karmakar2013capacity}, serves as a reference for assessing the performance achieved by the proposed approach. The case where there is no interference is also shown as an additional reference. Analyzing the results, it is evident that MADDPG with rate-splitting is very close to the upper bound. When compared with its no rate-splitting counterpart, we observe 2~dB gain when SNR is 12 dB or higher. MADDPG with RSMA also exhibits superior performance compared to MRT, ZF, and leakage-based precoding. MRT experiences challenges in the presence of severe interference, leading to a convergent behavior rather than maintaining an increasing average sum-rate curve. On the other hand, ZF, while immune to interference, struggles to achieve sufficiently high signal power, resulting in limited performance. Leakage-based precoding strikes a balance between desired signal power and leaking interference power, leading to higher rates than both MRT and ZF. 

The unique advantages of MADDPG with rate-splitting become apparent in the above comparison. Firstly, MADDPG utilizes the SINR as a metric, which is a more relevant than SLNR or SNR. Secondly, the incorporation of rate-splitting enables MADDPG to intelligently manage interference. In scenarios with weak interference, more power is allocated to private messages, while in the presence of strong interference, common messages are transmitted with higher power. The consideration of all possible decoding orders further enhances the adaptability of MADDPG with rate-splitting, resulting in superior performance compared to benchmark schemes. This comprehensive analysis sheds light on the nuanced benefits and capabilities of the proposed approach in addressing interference challenges in multi-antenna scenarios.

\begin{figure}[t]
 \centering
 \includegraphics[width=90mm]{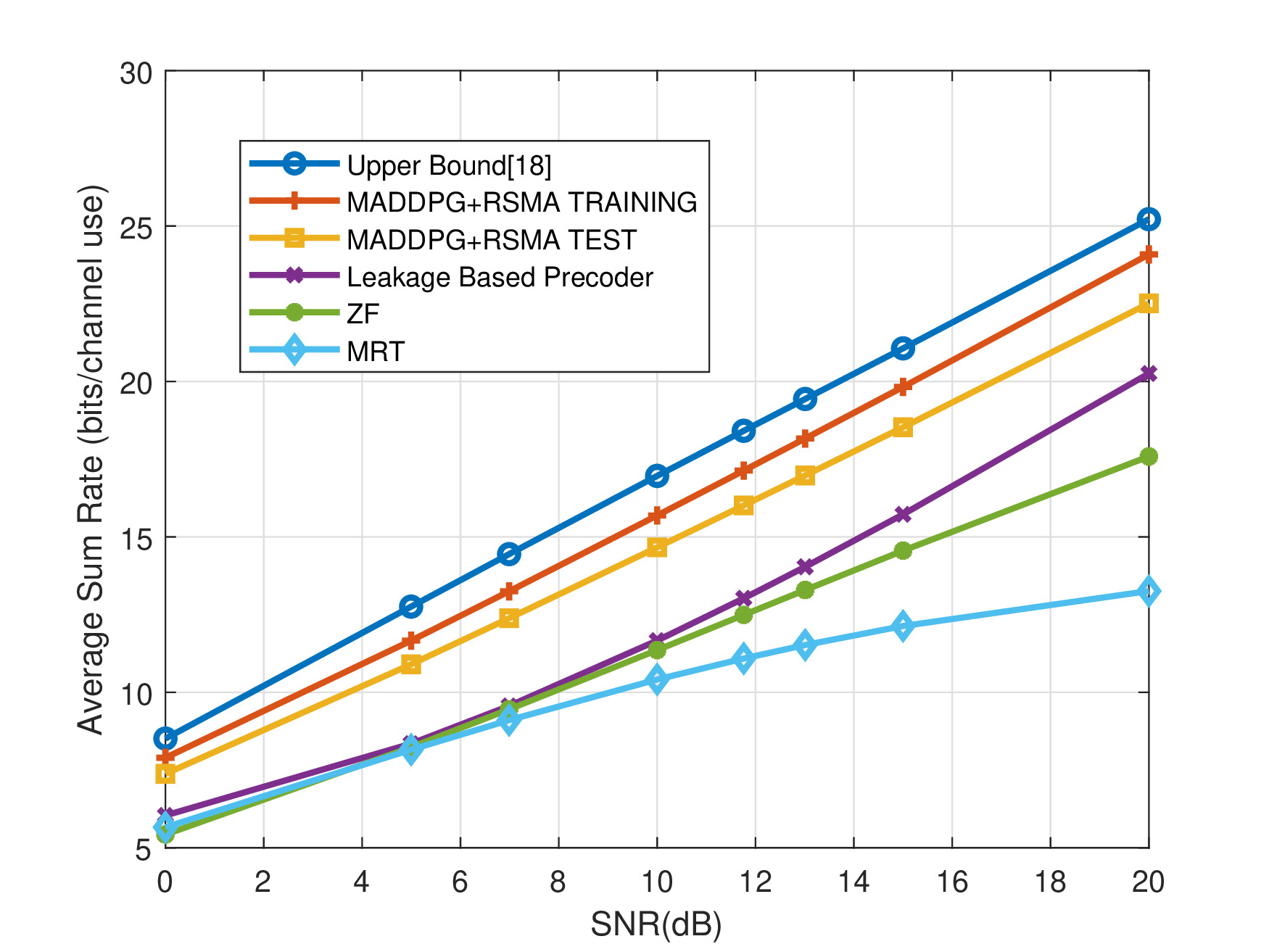}
\caption{ Average sum-rate achieved by MADDPG and the benchmark schemes for $M=3$, and $Q=3$. The MADDPG curves are obtained by averaging 5 runs, each having 200 time steps after the algorithm achieves convergence.}
 \label{Figure:mimo_rate_region}
\end{figure}

In Fig.~\ref{Figure:mimo_rate_region} we present a detailed analysis of the average sum-rate results for $M = 3$ and $Q = 3$. Our proposed MADDPG with RSMA method consistently outperforms benchmark schemes  ZF, MRT, and leakage-based precoder. This superiority can be attributed to the unique advantages offered by RSMA. By intelligently allocating power and managing interference, RSMA ensures more efficient resource utilization, resulting in enhanced overall system performance. Unlike in SISO, we also observe a gap between the training and test performance for MADDPG with RSMA. This increasing gap can be attributed to the growing number of parameters that need to be estimated as we move to more complex antenna configurations. Additionally, when compared with Fig.~\ref{Figure:single_antenna} and Fig.~\ref{Figure:multi_antenna} the gap between the upper bound and the MADDPG with RSMA is wider. However, we note that this gap can occur if the upper bound in \cite{karmakar2013capacity} is not tight for increasing number of transmit or receive antennas.

It is also important to note that, examining the number of training episodes across SISO, MISO, and MIMO configurations, a consistent upward trend was observed. This escalation in training episodes is a consequence of the heightened complexity inherent in transitioning from SISO to MIMO configurations. The increase in the number of antennas and parameters in the communication system demands more extensive training to optimize the reinforcement learning model effectively. The complexity of MIMO scenarios introduces additional intricacies, requiring the learning agent to navigate a larger action space and gain more information pertaining to the environment. Consequently, the iterative learning process is extended to accommodate the increased complexity and ensure the convergence of the reinforcement learning algorithm to an optimal policy. Therefore, the robustness and adaptability of our proposed method in the face of increasing complexity and interference make it a promising solution for advanced communication scenarios. %The inherent challenges of training a model with an expanding parameter space contribute to this performance gap.

\begin{figure}[t]
    \centering
    \includegraphics[width=90mm]{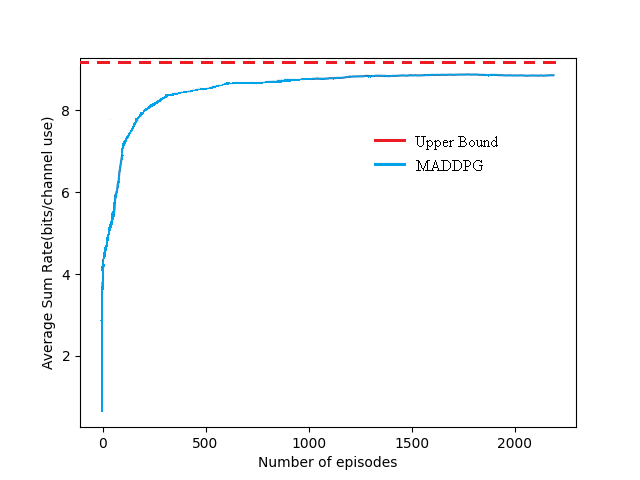}
    \caption{Convergence curve achieved by MADDPG for $M = 3$ and $Q=1$ when $SNR=10$ dB. The convergence curve is obtained by averaging 50 runs, each having 1000 time steps given the number of training episodes.}
     \label{convergence_miso}
\end{figure}

To examine the convergence behavior of the proposed algorithm in the MISO case with $M=3$ and $Q=1$, in Fig.~\ref{convergence_miso}, we present the average sum-rate versus the number of training episodes. For comparison, we include the upper bound\cite{karmakar2013capacity} as well. The convergence curve depict the performance evolution of the MADDPG algorithm, providing a visual representation of its learning trajectory over the course of training episodes. This visual analysis not only facilitates an assessment of the algorithm's convergence speed but also provides insights into how well the proposed solution converges towards the upper bound. %, shedding light on the effectiveness of the MADDPG algorithm in addressing the challenges posed by the MISO configuration. The convergence curve was drawn only for the MISO case, as a deliberate decision to avoid overwhelming the context with an excess of convergence curves and to maintain clarity and focus in the analysis.

\begin{figure}[t]
    \centering
    \includegraphics[width=90mm]{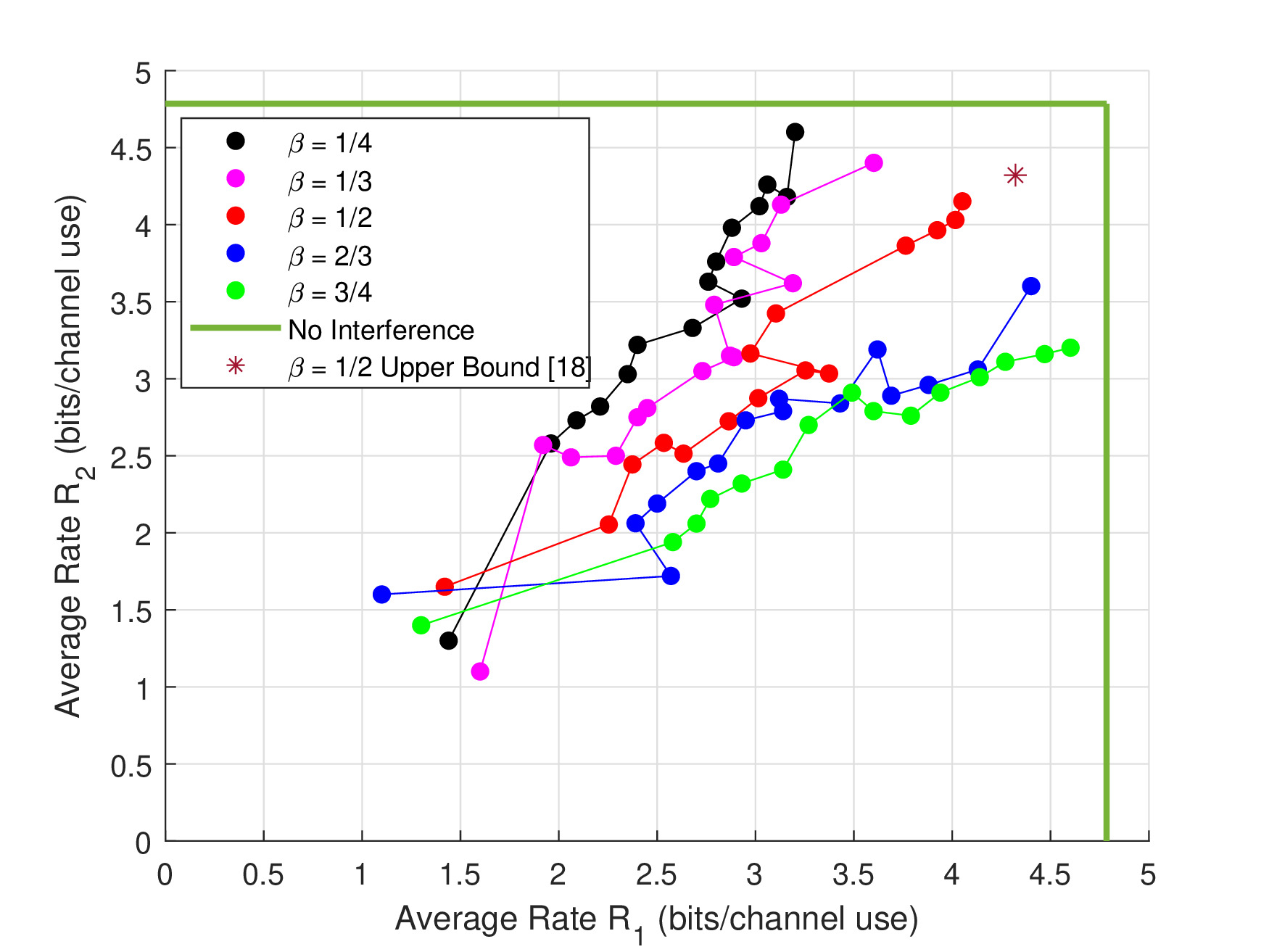}
    \caption{The evolution of the rate pair $(R_1, R_2)$ during training episodes for MADDPG with rate-splitting when $M = 3$, $Q=1$, $SNR=10$ dB and different values of $\beta$ as defined in \eqref{eq:maxrateobj}. Each dot represents 100 episodes of training, with the first dot appearing after 500 episodes. Subsequent dots are plotted every 100 episodes.}
    %\caption{Evolution of the weighted sum-rate for MADDPG with rate-splitting for $M = 3$ and $Q=1$ when $SNR=10$ dB and for varying $\beta$ defined in \eqref{eq:maxrateobj}. The skip from one dot to the next represents 100 episodes of training, with the dots appearing after a delay of 500 episodes.}
     \label{learning_curve_miso}
\end{figure}

In Fig.~\ref{learning_curve_miso}, we illustrate the evolution the rate pair $(R_1, R_2)$ as the number of episodes increases. This is examined under 10 dB SNR for $M=3$ and $Q=1$ for different weight parameter values; i.e. $\beta$ defined in \eqref{eq:maxrateobj}. In the figure, the skip from one dot to the next represents 100 episodes of training, with dots appearing after a delay of 500 episodes. The training process involves averaging our model over 100 runs, each comprising 1000 time steps realizations. The green outer box corresponds to the trivial outer bound with no interference, where each user can attain its individual capacity when there is no interference. We also include the rate pair point marked with a star to illustrate a point on the upper bound of the capacity region as defined by \cite{karmakar2013capacity}. The learning curve evolution analysis for different $\beta$ values allows us to explore how the rate region is obtained during the learning process. By observing how the system's performance changes with different $\beta$ values, we gain insights into the algorithm's behavior across a range of rate configurations, aiding in the assessment of its adaptability and responsiveness to changes in user rate priorities.

\begin{figure}[t]
 \centering
 \includegraphics[width=90mm]{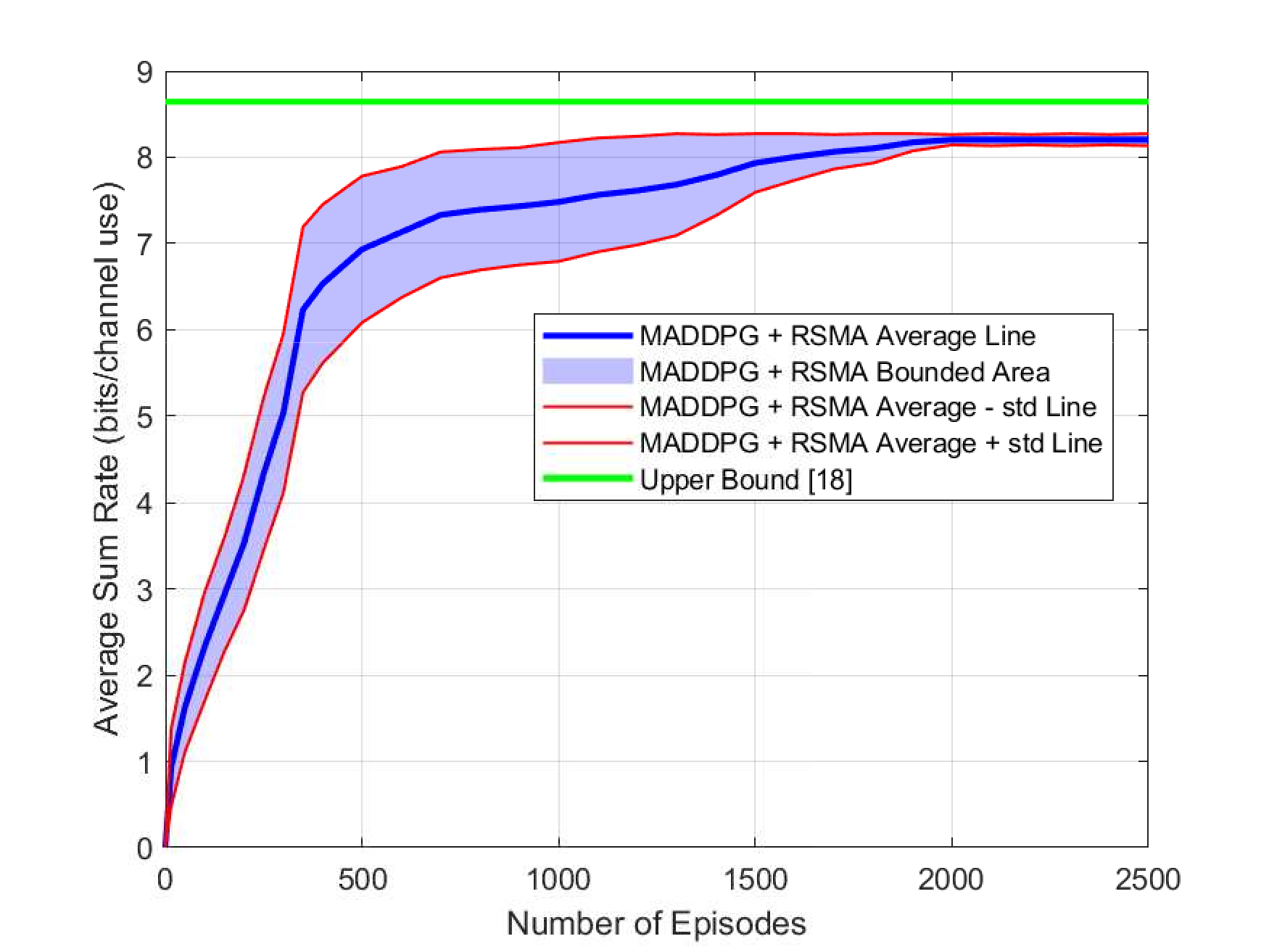}
 \caption{Confidence bound achieved by MADDPG for $M=3$ and $Q=1$. The confidence interval are obtained by averaging 1000 runs, each having 1000 time steps given the number of training episodes.}
 %Confidence bound achieved by MADDPG for $M=3$ and $Q=1$. The confidence interval are obtained by averaging 1000 runs, each having 1000 time steps after the algorithm achieves convergence.
 \label{Figure:confidence_bound}
\end{figure}

The confidence bounds of the reinforcement learning results are illustrated in Fig.~\ref{Figure:confidence_bound}. The figure, specifically obtained for the MADDPG with rate-splitting, showcase the robustness and reliability of the proposed approach. In assessing the performance of an algorithm, the inclusion of confidence bounds adds a layer of statistical significance to the obtained results. %The confidence bounds were derived by leveraging the information gathered from multiple runs of the MADDPG algorithm. 
To calculate the confidence bounds, the results were averaged over 1000 runs, each comprising of 1000 time steps. Subsequently, the standard deviation of each run was computed. By adding and subtracting this standard deviation from the averaged line, a confidence bound was established. % offering insights into the variability and consistency of the RL algorithm's performance across different runs. 
%This approach to confidence bounds is particularly valuable in reinforcing the reliability of the observed trends and outcomes. 
Comparison with the upper bound further contextualizes the reinforcement learning results, displaying the algorithm's proximity to the theoretical performance limits. %The robustness of the confidence bounds, derived through careful statistical analysis, enhances the credibility of the reinforcement learning results, providing stakeholders and researchers with a comprehensive perspective on the algorithm's performance under varying conditions. The comprehensive assessment of the confidence bounds, as illustrated in Fig.~\ref{Figure:confidence_bound}, augments the overall reliability and robustness of the acquired results derived from the MADDPG algorithm. 
%The promising aspect of these confidence bounds becomes more pronounced when considering the extensive nature of the evaluation, encompassing a total of 1000 runs, each comprising 1000 time steps. 
We also observe that after 2500 episodes, the algorithm consistently produces similar results. This stability and consistency in the performance indicate the robustness of our proposed solution. The algorithm's ability to maintain comparable outcomes across multiple episodes highlights its reliability in handling the complexities of the MISO scenario. This robust behavior is crucial for ensuring dependable and consistent performance under varying conditions, contributing to its practical applicability and effectiveness in real-world settings.

\begin{figure}[t]
     \centering
     \includegraphics[width=90mm]{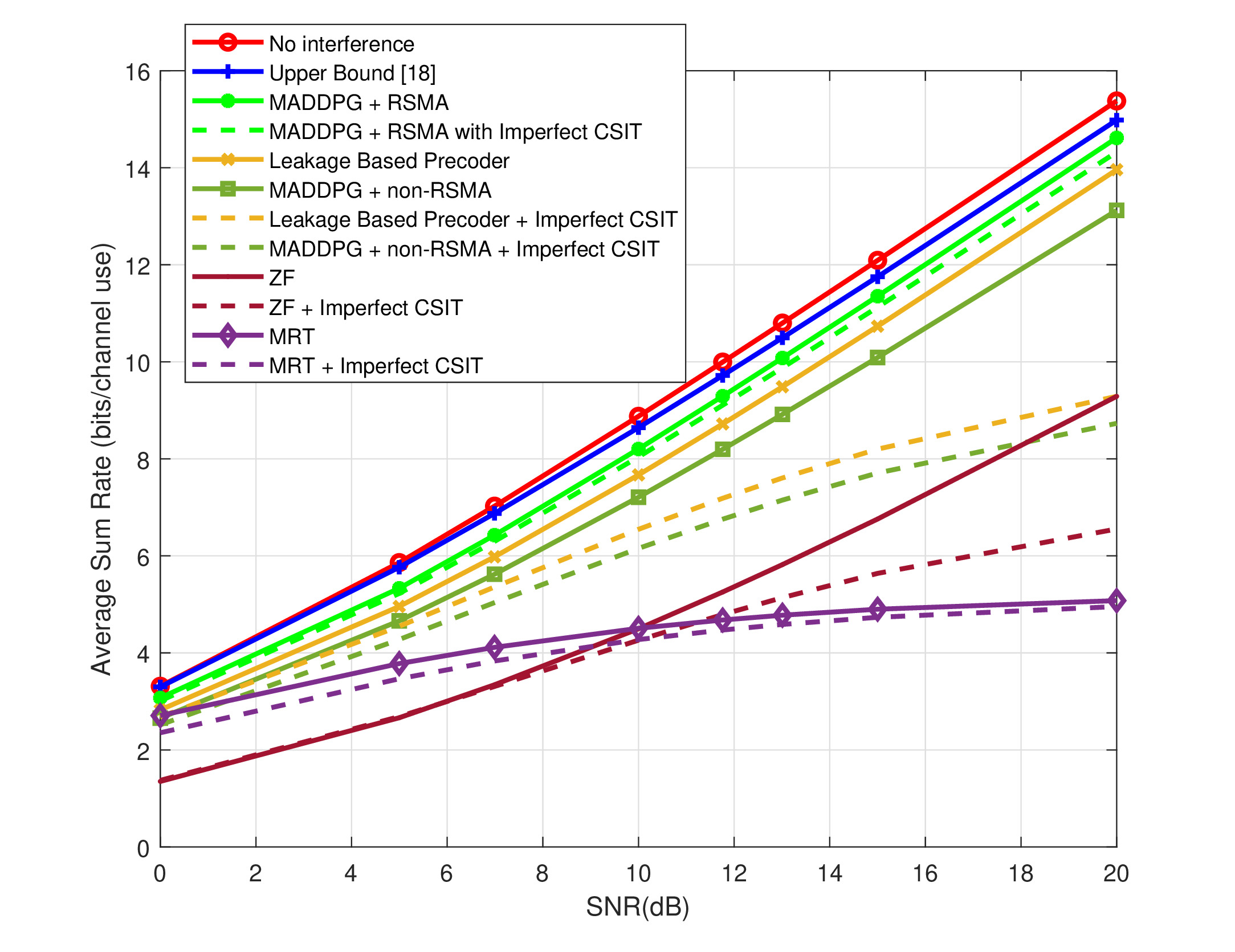}
     \caption{Effect of fixed channel estimation error on average sum rate versus SNR for $M=3$ and $Q=1$. The MADDPG curves are obtained by averaging 50 runs, each having 200 time steps after the algorithm achieves convergence.}
     \label{Figure:fixed_value_imp}
 \end{figure}

In Fig.~\ref{Figure:fixed_value_imp}, we systematically investigate the impact of a fixed value of imperfection over SNR, as defined in \eqref{channel_estimation}. The estimation error is set as $\frac{10^{-0.6}}{5}\mathcal{CN}(0, \sigma^2)$ with $\sigma^2 = 1$. In this analysis, the robustness of the proposed RSMA framework is highlighted. MADDPG with RSMA continues to follow the upper bound even under channel estimation errors. This quality becomes particularly pronounced when compared against alternative schemes, including ZF, MRT, leakage-based precoding, and non-rate-splitting MADDPG. In scenarios with imperfect channel state information at the transmitter (CSIT), the conventional schemes—ZF, MRT, leakage-based precoding, and non-rate-splitting MADDPG— encounter challenges in maintaining their performance due to their limited adaptability to variations in channel estimation errors. As channel estimation error variance is fixed with respect to SNR, the gap between perfect channel state knowledge and imperfect CSIT curves increase with increasing SNR for leakage based precoding, MADDPG with no rate-splitting and ZF. MRT relies on a straightforward strategy of amplifying the signals by the ratio of the received signal strength to the noise level, regardless of the specifics of the channel state. This approach minimizes the dependence on precise channel knowledge, making MRT less susceptible to the challenges associated with obtaining and utilizing accurate real-time channel state information. However, the simplicity that renders MRT resilient also contributes to its suboptimal nature. By adhering to a basic amplification strategy, MRT may not exploit the full potential of channel knowledge for optimizing transmission performance. However,  the unique rate-splitting mechanism in MADDPG exhibits exceptional proficiency in alleviating the deleterious effects of channel estimation errors and continues to perform very close to the upper bound. %The inherent resilience observed in RSMA translates into a robust and dependable performance, showcasing the superiority of the proposed framework under diverse SNR conditions. This comparative analysis reinforces the adaptability and reliability of RSMA, positioning it as a promising solution in realistic communication scenarios where channel imperfections are prevalent and challenging to address effectively.

\begin{figure}[t]
     \centering
     \includegraphics[width=90mm]{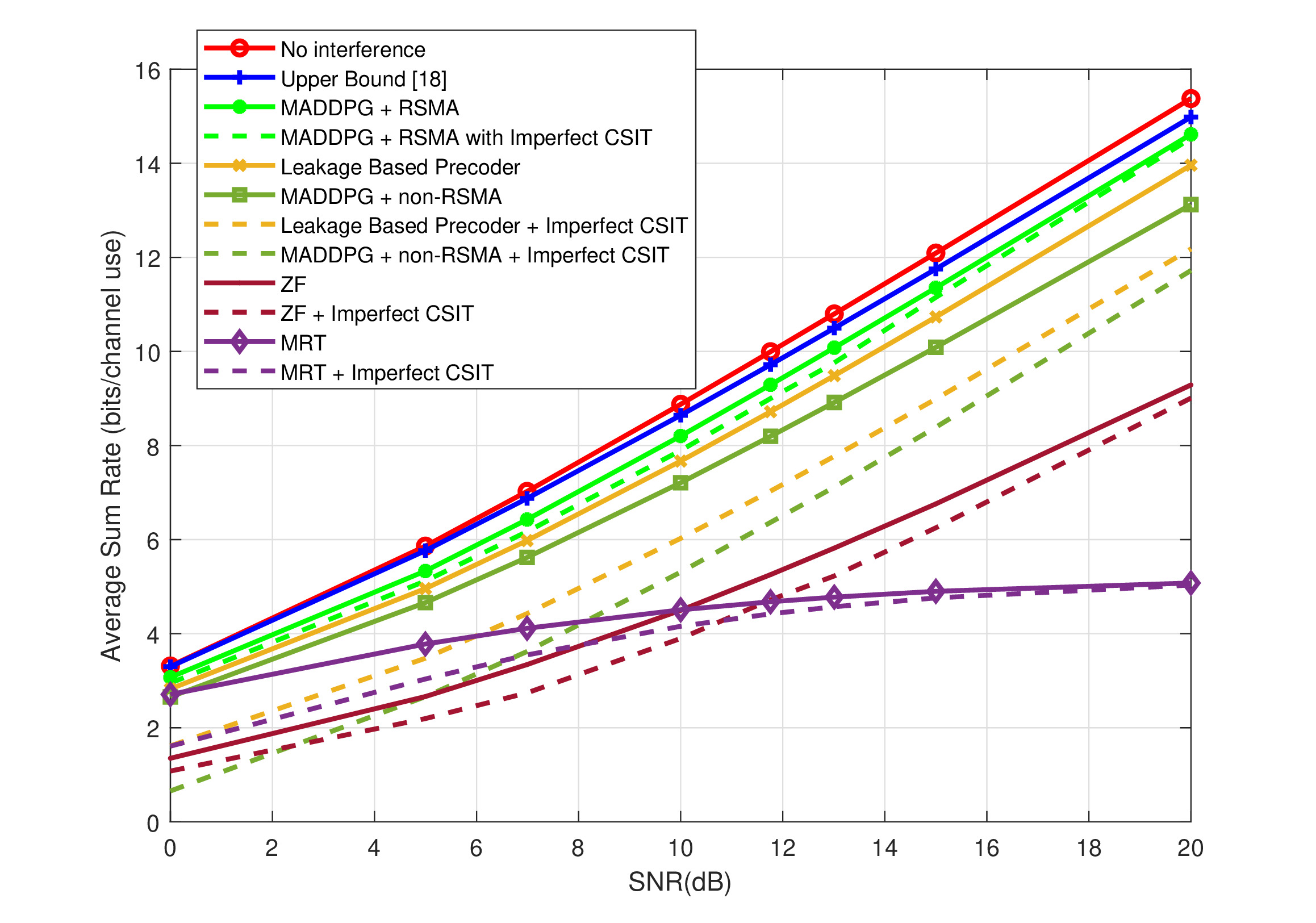}
     \caption{Effect of variable channel estimation error on average sum rate versus SNR for $M=3$ and $Q=1$. The MADDPG curves are obtained by averaging 50 runs, each having 200 time steps after the algorithm achieves convergence.}
\label{Figure:varying_value_imp}
 \end{figure}

Additionally, Fig.~\ref{Figure:varying_value_imp} explores the scenario where estimation error varies according to $\frac{SNR^{-0.6}}{5}\mathcal{CN}(0, \sigma^2)$, with $\sigma^2 = 1$. Although a fixed channel estimation error is a more limiting assumption, this scenario is more practical. If SNR is high, it is very likely that it is high both for the channel estimation phase and for the data transmission phase. We observe that MADDPG with rate-splitting, ZF and MRT all converge with their imperfect CSIT versions for high SNR. MADDPG with no rate-splitting and leakage based precoder also converge but at very high SNR.  Both Figs.~\ref{Figure:fixed_value_imp} and \ref{Figure:varying_value_imp} confirm the inherent resilience presented by RSMA, showcasing the superiority of the proposed framework under diverse SNR and channel estimation error conditions.

\begin{figure}[t]
 \centering
 \includegraphics[width=90mm]{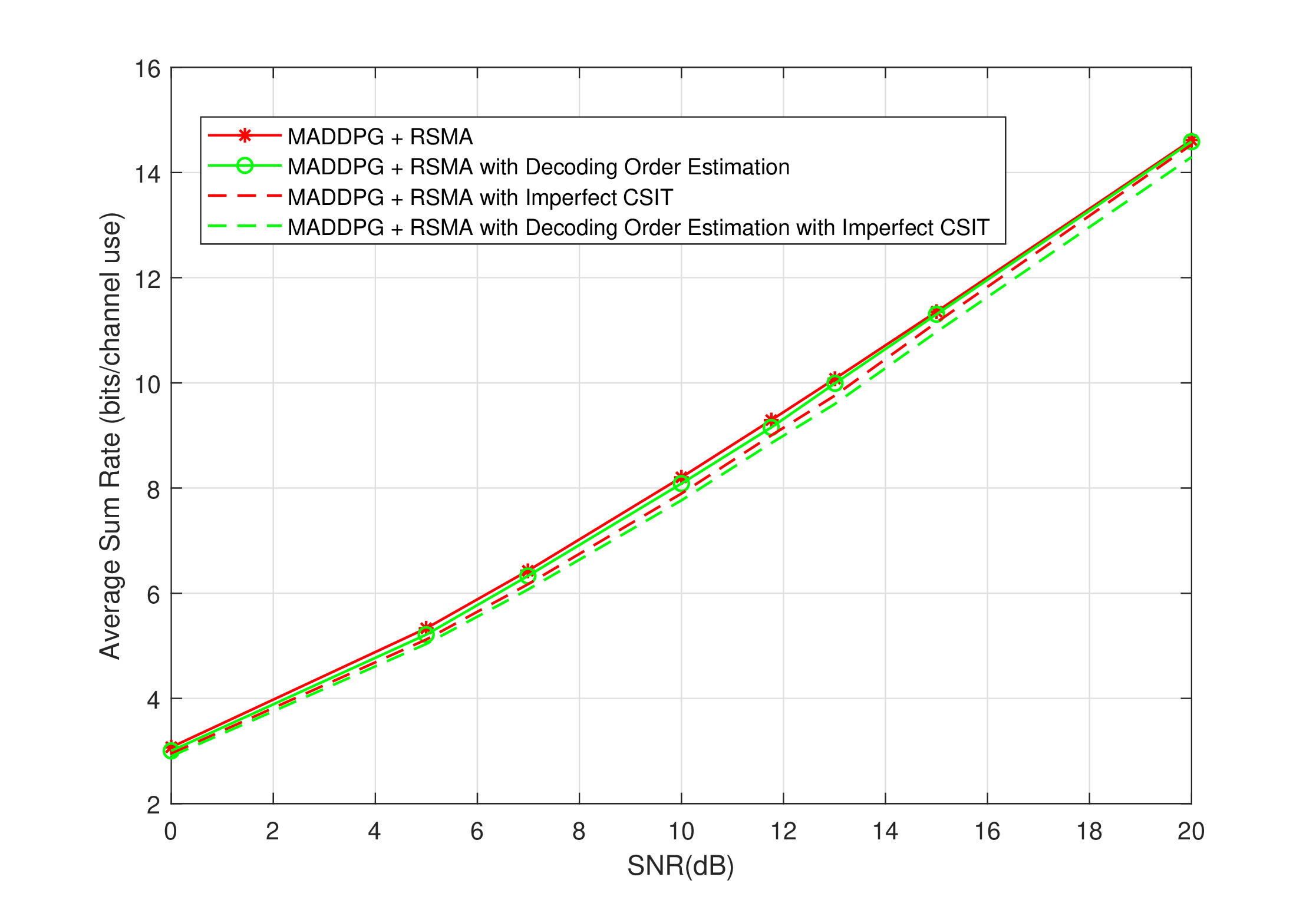}
 \caption{Average sum rate for MADDPG with decoding order estimation for $M=3$ and $Q=1$. The MADDPG curves are obtained by averaging 50 runs, each having 200 time steps after the algorithm achieves convergence.}
\label{Figure:decoding_order_estimation}
\end{figure}

In Fig.~\ref{Figure:decoding_order_estimation}, we study decoding order estimation. In the context of decoding order estimation, an additional actor network is introduced for each agent. Our investigation involves two key scenarios. Firstly, we examine the case where there is no channel estimation error. The second scenario considers the situation with a fixed value of channel estimation error over SNR, as defined earlier in this section. Remarkably, our findings reveal that decoding order estimation exhibits superior performance across all SNR values for both with and without channel estimation errors. This observation is particularly noteworthy as it shows the robustness and the efficacy of the proposed MADDPG framework. This result shows the ability of MADDPG to maintain superior performance even with decoding order estimation, addressing a crucial concern related to the complexity of RSMA, particularly in the SIC component. This suggests that by incorporating decoding order estimation, the system can achieve competitive performance while mitigating the associated complexity. %This nuanced exploration of decoding order estimation adds valuable insights to the understanding of RSMA's adaptability and efficacy in addressing practical challenges in communication systems.

%These insightful findings regarding the superior performance of decoding order estimation, particularly in the presence of fixed channel estimation errors and across various SNR regimes, are visually depicted in Fig.~\ref{Figure:decoding_order_estimation}. The depicted results provide a clear and illustrative representation of the robustness and effectiveness of the proposed MADDPG framework in addressing the complexities associated with decoding order estimation in the context of the Rate-Splitting Multiple Access (RSMA) structure.

\section{Conclusion and Future Work}\label{sec:conclusion}
In the context of advancing communication systems from 5G to 6G and beyond, rate-splitting multiple access has emerged as a notable transmission strategy. In rate-splitting, data streams are divided into common and private parts to strike the balance between treating interference as noise and decoding all unintended messages. However, especially in multiple antenna scenarios, finding the optimal precoders and transmit power levels for rate-splitting for all common and private data streams happens to be quite complex. To alleviate this issue, in this paper we propose computing the optimal precoders and transmit power levels for rate-splitting with deep reinforcement learning. Specifically, we use multi-agent deep deterministic policy gradient framework, in which decentralized agents with limited information collectively learn from a centralized critic to optimize actions in a multi-dimensional continuous policy space.
This framework allows for centralized learning while enabling decentralized execution and offers a scalable and decentralized solution for interference management in multiple antenna environments. Simulation results demonstrate the effectiveness of our proposed rate-splitting method, achieving information-theoretical sum-rate upper bounds in single-antenna scenarios and impressive proximity to the upper bounds in multiple-antenna scenarios. Additionally, our method outperforms alternative approaches, zero-forcing, leakage-based precoding, and maximal ratio transmission, showcasing superior weighted sum-rate performance in both single and multiple antenna cases.

Moreover, our work enhances the learning algorithm by addressing channel estimation errors and optimizing decoding order of common and private streams. Both aspects are important for prospective practical applications for improved robustness and efficiency. Especially, the latter is overlooked in the literature and generally a fixed decoding order is assumed at the receivers. Comparative analyses highlight the superiority of our approach, particularly in determining decoding orders and showing resilience against channel estimation errors. The investigation of learning curve evolution, variable weights in weighted sum rate calculation and confidence bounds complement our study for a  comprehensive treatment of multi-agent deep deterministic policy gradient algorithm for rate-splitting. 

The proposed MADDPG framework could be easily extended to handle scenarios with multiple users and to design new algorithms that take energy efficiency into account. Also, exploring hierarchical reinforcement learning approaches to tackle the complexity of rate-splitting optimization, and investigating meta-learning techniques to reduce the need for extensive training in new scenarios are of interest. In short, the future of using deep reinforcement learning for rate-splitting in wireless communications holds great potential for optimizing resource allocation, improving spectral and/or energy efficiency, and enabling more dynamic and adaptable wireless networks. Our work is a first step towards addressing these challenges and will lead to significant advancements in wireless communication systems.

\bibliographystyle{ieeetr} % Replace 'plain' with your desired style
\bibliography{references} % Replace 'references' with the name of your .bib file without the extension

\end{document}